\documentclass[11pt]{article}

\usepackage{url}
\usepackage{amsmath,amssymb}           
\usepackage{amscd}                     
\usepackage{epsfig}                    
\usepackage[matrix,arrow]{xy}          
\usepackage{xspace}                    
\usepackage{stmaryrd}                  
\usepackage{slashed}
\usepackage{mathtools}
\newcommand{\fsl}[1]{\ensuremath{\mathrlap{\!\not{\phantom{#1}}}#1}}

\usepackage{jheppub}                   

\newcommand{\bq}{\begin{equation}}
\newcommand{\eq}{\end{equation}}
\newcommand{\bea}{\begin{eqnarray}}
\newcommand{\eea}{\end{eqnarray}}

\newcommand{\dd}{\mathrm{d}}

\newcommand{\ii}{\mathrm{i}}

\newcommand{\bbR}{\mathbb{R}}

\DeclareMathOperator{\SO}{\mathit{SO}}

\DeclareMathOperator{\Cliff}{Cliff}

\DeclareMathOperator{\tr}{tr}

\DeclareMathOperator{\vol}{vol}

\newcommand{\ph}[1]{\phantom{#1}}

\newcommand{\Dgen}{{D}}

\newcommand{\LC}{\nabla}

\newcommand{\Ric}{\mathcal{R}}
\newcommand{\Scalar}{\mathcal{R}}

\DeclareMathOperator{\Edd}{\mathit{E_{d(d)}}}

\newcommand\ft[2]{{\textstyle\frac{#1}{#2}}}
\mathchardef\mhyphen="2D

\newcommand{\sla}{\slash\!\!\!\!}
\newcommand{\LL}{\mbox{Lichnerowicz}}

\begin{document}
\begin{titlepage}

\begin{center}

\rightline{\small IPhT-t17/079}

\vskip 1.0cm

{\Large \bf M-theoretic Lichnerowicz formula}
\vskip 0.7cm
{\Large \bf
 and  supersymmetry 
}

\vskip 1.5cm

{\bf Andr\'e Coimbra}$^a$ and  {\bf Ruben Minasian}$^{b}$\\

\vskip 0.5cm

{}$^{a}${\em Institut des Hautes Etudes Scientifiques  \\
 F-91440 Bures-sur-Yvette, France }

\vskip 0.4cm

{}$^{b}${\em Institut de Physique Th\'eorique, Universit\'e Paris Saclay,  CNRS, CEA \\
F-91191 Gif-sur-Yvette, France }

\vskip 0.7cm

\end{center}

\vskip 1.0cm

\begin{center} {\bf ABSTRACT } \end{center}
A suitable generalisation of the Lichnerowicz formula can relate the squares of supersymmetric operators to the effective action, the Bianchi identities for fluxes, and some equations of motion. Recently, such formulae have also been shown to underlie the (generalised) geometry of supersymmetric theories. 
 In this paper, we derive an M-theoretic Lichnerowicz formula that describes eleven-dimensional supergravity together with its higher-derivative couplings. The first corrections to the action appear at eight-derivative level, and the construction yields two different supersymmetric invariants, each  with a free coefficient.  We discuss the restriction of our construction to seven-dimensional internal spaces, and implications for compactifications on manifolds of $G_2$ holonomy. Inclusion of fluxes and computation of contributions with higher than eight derivatives are also discussed.

\vfill

\noindent\today


\end{titlepage}


\tableofcontents

\section{Introduction}

Recent development of generalised complex geometry \cite{Hitchin, Gualtieri} has been intimately related with our understanding of supersymmetric theories and has provided a natural language for classifying supersymmetric vacua and revealing the structure of off-shell supersymmetry in effective theories~\cite{GMPT,
Coimbra:2014uxa}. 
It is still, however, an open question whether  generalised complex geometry can capture the structure of stringy perturbative corrections to the effective theories. If so one may hope that the geometric insight might one day translate into computational tools.

In its most basic formulation, GCG unites the geometric data with the antisymmetric two form field and puts diffeomorphisms and gerbe gauge transformations on equal footing~\cite{Hitchin}. This unified formalism allows a transfer of many results of Riemannian geometry to the generalised tangent bundle~\cite{Gualtieri}. Notably, a generalised vanishing torsion condition on a generalised metric-connection yields a notion of a generalised Ricci tensor which fully captures the physics of the ten-dimensional NS sector -- the action and the equations of motion~\cite{CSW1}. Interestingly, and crucially for this work, the part covering only the metric and two-form field can alternatively be derived by using a version of the Lichnerowicz formula  with closed three-form torsion $H$, due to Bismut. The standard Lichnerowicz formula~\cite{Lichnerowicz} for a Levi-Civita connection states that the difference of the squares of the covariant derivative and the Dirac operator acts tensorially on a spinor and can serve as  a definition of a scalar curvature $\mathcal R$:
\begin{equation}
\label{eq:LC_Lich}
     ( (\sla \nabla)^2  -  \nabla^a \nabla_a)\varepsilon
          = - \tfrac14 \mathcal R \varepsilon   .
 \end{equation}
Here and throughout the paper $\nabla$ denotes the covariant derivative with respect to the torsion-free Levi-Civita connection. 
The Bismut version~\cite{Bismut} finds a pair of operators with torsion such that the difference of their squares is again tensorial
\begin{equation}
\label{eq:Bismut}
    (\sla \nabla^{\scriptscriptstyle H})^2 \varepsilon -  (\tilde{\nabla}^{\scriptscriptstyle H})^a \tilde\nabla^{\scriptscriptstyle H}_a  \varepsilon
          = - \tfrac14 \mathcal R \varepsilon  +\tfrac{1}{48} H^2\varepsilon   ,
 \end{equation}
 where
\begin{equation}
\begin{aligned}
\label{eq:Bismut2}
     \tilde{\nabla}^{\scriptscriptstyle H}_a\varepsilon
          &=  \nabla \epsilon  +\tfrac18 H_{abc}\gamma^{bc}\varepsilon ,\\
     \sla \nabla^{\scriptscriptstyle H}\varepsilon
          &=  \sla\nabla \varepsilon  +\tfrac{1}{24} H_{abc}\gamma^{abc}\varepsilon .\\
\end{aligned}
 \end{equation}
Note that the Dirac operator is no longer the trace of the covariant derivative with torsionful connection! Also, strictly speaking, the tensor has two parts now -- a scalar component given by a sum of the scalar curvature and $H^2$ and a four-form given by the (vanishing) exterior derivative of $H$.

Interestingly enough it is the the inclusion of the dilaton that requires a truly generalised treatment. The key here is having a different action of the same operator on a spinor and a vector-spinor. The above-mentioned pair of operators is then seen as a component of the same ``generalised Levi-Civita connection'' $D$. This pair also appears in the respective supersymmetry variations of gravitino and dilatino 
\begin{equation}
\label{eq:FermVar2}
\begin{aligned}
   \delta\psi_a &= D_a \varepsilon
      = \nabla_a\epsilon + \tfrac{1}{8}H_{a\bar{b}\bar{c}}
           \gamma^{\bar{b}\bar{c}} \varepsilon , \\
   \delta\rho &= \gamma^{\bar{a}}D_{\bar{a}}\varepsilon
     = \gamma^{\bar{a}}\nabla_{\bar{a}} \varepsilon
        + \tfrac{1}{24}H_{\bar{a}\bar{b}\bar{c}}
          \gamma^{\bar{a}\bar{b}\bar{c}}\varepsilon
        -\gamma^{\bar{a}}(\partial_{\bar{a}}\phi)\varepsilon ,
\end{aligned}
\end{equation}
and the Lichnerowicz formula yields:
\begin{equation}
\label{eq:newBismut*}
     ( (\sla D)^2  - D^aD_a) \varepsilon
          =  -\tfrac14 S \varepsilon - \gamma^{abcd} I_{abcd} \varepsilon ,
 \end{equation}
where on the right hand side we find  the bosonic NS action $S$, and the Bianchi identity $I_{abcd} = \nabla_{[a} H_{bcd]}$.

The passage to non-closed torsion $H$ requires an extension of the generalised tangent bundle and the inclusion of gauge fields, eventually leading to a generalised complex description of heterotic strings.  Once one properly extends the notion of covariant derivative to also cover  gaugino variations, the Lichnerowicz theorem yields the ten-dimensional $N=1$ supergravity coupled to supersymmetric YM, and can further accommodate the heterotic Bianchi identity and the higher derivative terms in the effective action that serve its supersymmetry completion~\cite{Coimbra:2014qaa}. 
Interestingly, a single four-derivative correction to the Bianchi identity requires an infinity of terms in the supersymmetry variations and hence in the effective action \cite{Bergshoeff:1988nn, Bergshoeff:1989de}.
Using the simple principle of building quadratic combinations of first order operators appearing in the supersymmetry variations that act only tensorially, allows one to compute the $\mathcal{O}(\alpha'^4)$ \cite{in progress}.

It is also possible to fully geometrise both the NS and RR sector, or alternatively describe M-theory, of the internal $d$-dimensional space of a generic supersymmetric compactification  in the language of exceptional generalised geometry, where the generalised tangent bundle corresponds to a representation of an exceptional $\Edd$ group~\cite{chris,PW}. The bosonic action can again be written in the form of a Lichnerowicz equation for generalised Levi--Civita connections~\cite{Coimbra:2012af}. Schematically
\begin{equation}
\label{eq:EddLich}
     D\otimes_S D\otimes_S \varepsilon   - D \otimes_S D \otimes_J  \varepsilon 
          =  - \tfrac14 S_{\text{B}} \varepsilon ,
 \end{equation}
where $\otimes_S,\otimes_J$ denote projections to the generalised spinor bundles into which, respectively, the spinor and the gravitino embed, and now $S_{\text{B}}$ is the full bosonic action, with RR fields, of the dimensionally restricted theory.

 At this point one would like to dream that the suitable versions of the Lichnerowicz formula underlie every supersymmetric theory. Eleven-dimensional supergravity, even at a classical level, would appear to be the best candidate for undermining such a belief. Indeed the theory does not have a dilatino, and hence there is no natural Dirac operator other than the trace of the gravitino variation. Moreover, any Lichnerowicz type formula like 
\eqref{eq:newBismut*} built out of operators linear in fields would yield an action that is at most quadratic. From other side in spite of its simplicity the eleven-dimensional supergravity action contains a cubic Chern-Simons interaction.
     
Yet, as we shall see, there is an appropriate generalisation. Indeed, using the Levi-Civita connection and the four-form flux $G$ one may find an operator $ \nabla^G_a $ such that it yields a tensorial polyform $\rho$ (there are no derivatives acting on $\varepsilon$ on the right hand side) when squared:
\begin{equation}
\label{eq:11LL}
\gamma^{ab} {\nabla}_a^G\nabla_b^G\varepsilon  = -\tfrac14 \rho \cdot \varepsilon  .
\end{equation}
Moreover requiring that $\rho$ has only an 11-form (or scalar) and an 8-form component\footnote{We are dropping here the five-form which correspond to the Bianchi  identity $\nabla_a G_{bcde} \gamma^{abcde}$.} uniquely fixes   $ \nabla^G_a $  to be the operator that appears in the gravitino variation $ \delta \psi_a =  \nabla^G_a \varepsilon$. Furthermore, we may observe that up to integration by parts the Mukai pairing 
\begin{equation}
\label{eq:11La}
\mathcal{L}_B = < 1 +  C, \rho>  =  \rho|_{11} - C \wedge \rho|_{8} ,
\end{equation}
gives the Lagrangian for eleven-dimensional supergravity.

Note that we have used very little here. As far as the generalised geometry constructions go, we seem to just pass to what locally looks like $TM \oplus \Lambda^2 T^*M$, hence avoiding the complications of exceptional  (and possibly infinite-dimensional) geometries, dual gravitons, need to stick to linearised equations and so on. Some or all of these issues arising in the full geometrisation of M-theory and U-dualiy groups will eventually have to be faced, and contrasted to the simplicity of \eqref{eq:11La}. Instead of doing all this, we shall try to build on this construction, and modify  $ \nabla^G_a $ by higher-derivative contributions and verify if the method of requiring bilinears with tensorial action can reproduce the  quantum higher-derivative corrections to M-theory.

The formula \eqref{eq:11LL}, where now the corrected operator $ \nabla^G_a $ will  a priori be all order in derivatives, is what we shall call M-theoretic $\LL$ formula. Actually we shall see that the first modifications from the the one-derivative supergravity operator will start at seven derivatives, and the ``minimal" version of the corrected operator can contain only $6l + 1$ derivatives (for integer $l$). When such operator is plugged into \eqref{eq:11La} it will clearly produce a series of terms in $\mathcal{L}_B$ with $2\times(3l+1)$ derivatives.\footnote{Note that the opposite is not true. An action with   $2\times(3l+1)$ derivatives does not necessarily restrict  ${\nabla}_a^G$ to have  {\sl only}  $(6l+1)$-derivative terms.} The use of the computer algebra system Cadabra~\cite{Cadabra} was crucial to verify these computations.

The structure of the paper is as follows. In section \ref{sec:11sugra}, we review the eleven dimensional supergravity and the first $\sim R^4$ quantum corrections. The M-theoretic Lichnerowicz formula is set up in section \ref{sec:M-Lich}. Section \ref{sec:high} extends the construction to higher derivative terms, verifying that the first corrections appear at eight-derivative level, and constructing explicitly two different supersymmetric invariants (each coming with a free coefficient). We specialise on the case of seven-dimensional internal spaces in section \ref{sec:7D}. Possible extensions of our construction are discussed in section \ref{sec:future}.

\section{Eleven-dimensional supergravity}
\label{sec:11sugra}

Let us start by reviewing classical eleven-dimensional supergravity~\cite{Cremmer:1978km}, to 
leading order in the fermions, following the conventions
of~\cite{Gauntlett:2002fz,Coimbra:2012af}. The bosonic fields are the eleven-dimensional Lorentzian metric $g_{ab}$ and a three-form Abelian gauge potential $C_{abc}$ and there is a fermionic gravitino field $\psi_a$. 

Writing $\Scalar$ for the Ricci scalar for the Levi--Civita connection $\nabla$ and $G=\dd
C$ for the four-form field-strength, the bosonic action is  given by  
\begin{equation}
\label{eq:NSaction}
   S_{\text{B}} = \frac{1}{2\kappa^2}\int \Big(
       \Scalar \ast 1 - \tfrac{1}{2}G\wedge *G
          - \tfrac{1}{6}C\wedge G\wedge
          G 
       \Big) ,
\end{equation}
and taking $\gamma^a$ to be the $\Cliff(10,1;\bbR)$ gamma matrices, the
fermionic action is
\begin{equation}
\begin{aligned}
   S_{\text{F}} &= \frac{1}{\kappa^2} \int \sqrt{-g} 
       \Big( \bar\psi_a \gamma^{abc} \LC_b \psi_c 
           + \tfrac{1}{96} G_{a_1\dots a_4}
              \bar\psi_b ( \gamma^{a_1\dots a_4bc} +12  \gamma^{a_1a_2}g^{a_3 b}g^{a_4 c})\psi_{c} 
		\Big) .
\end{aligned}
\end{equation}

This leads to the equations of motion for the metric and gauge field
\begin{equation}
\label{eq:eom11}
\begin{aligned}
  \Ric_{ab} - \tfrac{1}{12} \left( 
           G_{ac_1c_2c_3} 
           G_b{}^{c_1c_2c_3}
		- \tfrac{1}{12} g_{ab} G^2 \right)
      &= 0 , \\
   \dd * G + \tfrac{1}{2} G\wedge G &= 0 , 
\end{aligned}
\end{equation}
where $\Ric_{ab}$ is the Ricci tensor, and the gravitino equation of motion is
\begin{equation}
  \gamma^{abc} \LC_b \psi_c + \tfrac{1}{96} \left(
      G_{c_1\dots c_4} \gamma^{abc_1\dots c_4} 
      + 12 G^{ab}{}_{c_1c_2} 
         \gamma^{c_1c_2} \right) \psi_b = 0 .
\end{equation}

This action is invariant under supersymmetry, with the transformation of the bosons given by
\begin{equation}
\begin{aligned}
   \delta g_{ab} &= 2 \bar\varepsilon \gamma_{(a} \psi_{b)} ,\\
   \delta C_{abc} 
      &= -3 \bar\varepsilon \gamma_{[ab} \psi_{c]} ,
\end{aligned}
\end{equation}
while the supersymmetry variation of the gravitino is
\begin{equation}
\label{eq:grav-var}
  \delta \psi_a = \LC_a \varepsilon 
     + \tfrac{1}{288} \left(\gamma_a{}^{b_1\dots b_4}
        - 8 \delta_a{}^{b_1} \gamma^{b_2b_3b_4} 
        \right) G_{b_1\dots b_4} \varepsilon ,
\end{equation}
where $\varepsilon$ is the supersymmetry parameter.

\subsection{Review of higher-derivative terms}
\label{sec:11d-rev}

In presence of M5 branes, this classical action is not consistent without the inclusion of certain higher-derivative terms. Indeed, since a single M5 supports a chiral six-dimensional $(2,0)$ tensor multiplet on the worldvolume, it is anomalous. The anomaly is a descendant of a particular eight-derivative eight-form polynomial. To cancel this anomaly via the inflow mechanism, one needs bulk couplinsg which in absence of M5 sources are invariant under diffeomorphisms and $C$-field gauge transformations, but become anomalous in the presence of the branes. Moreover this anomaly should restrict to the six-dimensional worldvolume. 

The anomaly cancelling bulk term is of course only a part of the story, as it is hardly supersymmetric on its own. From one side it should receive contributions from $G$-flux. From other, it requires completion and other (non-anomalous) couplings  with the same number of derivatives. Furthermore, since M-theory is supposed to be the strong coupling limit of type IIA strings, the eleven-dimensional couplings should also be seen in the decompactification limit of the string theory.

The eight-derivative couplings in string theory have a long history and (without fluxes) have been computed using string perturbation theory. In IIA there are only tree-level and one loop terms, while IIB also has D-instanton contributions. The tree level corrections are the same in both theories and take the form \cite{Gross:1986iv, Grisaru:1986dk}
\begin{equation}\label{eq:tree}
e^{-1}\mathcal L\sim e^{-2\phi}(t_8t_8R^4 - \ft14 E_8) \, ,
\end{equation}
and the CP-even sector of the one-loop is given by \cite{Sakai:1986bi}
\begin{equation}\label{eq:one-even}
e^{-1}\mathcal L_{\rm CP\mhyphen even}\sim (t_8t_8R^4\pm\ft14 E_8),
\end{equation}
where the top (bottom) sign is for the IIA (IIB) theory. The $E_8$ term can be written using two totally antisymmetric $\epsilon$-tensors.\footnote{We follow the $R^4$ conventions of \cite{Antoniadis:2003sw}  and define $$t_8 t_8 R^4 = t_{a_1\cdots a_8} t_{b_1\cdots b_8}
R^{a_1a_2}{}_{b_1b_2} R^{a_3a_4}{}_{b_3b_4}
R^{a_5a_6}{}_{b_5b_6}R^{a_7a_8}{}_{b_7b_8}, $$ and $$E_8 = 8! \times \delta^{b_1\cdots b_8} _{a_1 \cdots a_8} R^{a_1a_2}{}_{b_1b_2} R^{a_3a_4}{}_{b_3b_4}
R^{a_5a_6}{}_{b_5b_6}R^{a_7a_8}{}_{b_7b_8}.$$ Note that for any antisymmetric matrix $M$, $ t_8 M^4 =  24 \left(\tr M^4 - \ft14 (\tr M^2)^2\right),$ and there is a useful relation between two quantities given by $\ft14 E_8 = t_8 t_8 R^4  + 192 \,  \tr R_{a b} R_{c d}  \tr R_{a c} R_{b d} 
 - 768 \, \tr R^{a b c d} R_{a}\,^{e} \,_{c} \, ^{f} R_{e} \,^{g} \, _{b} \,^{h} R_{f g d h} + \mbox{Ricci terms}$.}
In addition, the IIA theory has a CP-odd
one-loop term  $\mathcal L_{\rm CP\mhyphen odd}\sim B_2\wedge[\tr R^4-\ft14(\tr R^2)^2]$ \cite{Vafa:1995fj, DLM}.
Its lift to eleven dimensions 
\begin{equation}
\label{eq:Bwed}
\mathcal L_{\rm 11d}\sim C_3\wedge[\tr R^4-\ft14(\tr R^2)^2] ,
\end{equation}
cancels the M5 anomalies via inflow. The one-loop CP-even terms \eqref{eq:one-even} also lift, while the tree-level contribution \eqref{eq:tree} is suppressed \cite{Green:1997di}.\footnote{The string $l$-loop terms surviving in the eleven-dimensional limit are $\sim R^{3l+1}$. Note that due to the relation to anomalies, the one loop term is not renormalised, and the terms with a number of  derivative higher than eight should not contain any top-form couplings of the type \eqref{eq:Bwed}.}

\medskip
\noindent
Our expectation is that the bosonic action of supersymmetric M-theory (just like any other supersymmetric theory) should be formulated as a kind of Lichnerowicz theorem. As a first step, we shall verify that this is the case for eleven-dimensional supergravity and recast the action as a Mukai transformed square of a first-order operator. Unsurprisingly, the operator in question turns out to be the covariant derivative with $G$-flux terms included that appears in the gravitino supersymmetry variation \eqref{eq:grav-var}. 

The modifications of this operator with higher-derivative terms included should in turn yield the supersymmetry variation for the M-theory action with higher-derivative terms included. Such supersymmetry modifications have been discussed e.g. in \cite{Hyakutake:2005rb, Soueres:2016qre}; seven-derivative corrections were computed explicitly in  \cite{Peeters:2000qj} by reading them off from the effective action. Unfortunately, these are written in a different basis from ours, and due to the complicated nature of the terms the direct comparison of two results in not straightforward. For special cases of seven and eight-dimensions (for manifolds of $G_2$ and $Spin(7)$ holonomy), simplified transformations have been suggested in \cite{Lu:2003ze}.

The potential usefulness of the explicit form of the transformations in a convenient basis for e.g. compactifications makes our computation worthwhile. Our main motivation has been to test the $\LL$ method, and the hypothesis that it should underlie any supersymmetric theory. We also hope that this will provide a better systematics for still open questions such as inclusion of fluxes (which is very hard even in string theory, due to the need to perform higher-point calculations), or higher derivative terms (which in string theory, for the terms that appear in eleven dimensions, would also correspond to going to higher string loops). Finally, establishing an M-theoretic $\LL$ formula, analogous to the string theoretic cousins that have a generalised geometric origin, will hopefully lead to new  insights about the geometry of M-theory.

\section{ Lichnerowicz formula for eleven-dimensional supergravity}
\label{sec:M-Lich}

Let us write the operator that appears in the gravitino variation as 
\begin{equation}
\label{eq:grav-var2}
  \delta \psi_a = \LC_a \varepsilon 
     + \tfrac{1}{288} \left(\gamma_a{}^{b_1\dots b_4}
        - 8 \delta_b{}^{b_1} \gamma^{b_2b_3b_4} 
        \right) G_{b_1\dots b_4} \varepsilon = \nabla^G_a \varepsilon .
\end{equation}
It has long been known (see for example~\cite{Gauntlett:2002fz}) that there is an ``integrability'' condition satisfied by this connection
\begin{equation}
\label{eq:integrability cond}
\gamma^a[\nabla^G_a,\nabla^G_b]\varepsilon \propto (\text{all bosonic eoms})_b\cdot\varepsilon .
\end{equation}

We can view this as being a consequence of supersymmetry. If we also denote the M-theory Rarita--Schwinger operator in the gravitino equation of motion, 
\begin{equation}
  \gamma^{abc} \LC_b \psi_c + \tfrac{1}{96} \left(
      G_{c_1\dots c_4} \gamma^{abc_1\dots c_4} 
      + 12 G^{ab}{}_{c_1c_2} 
         \gamma^{c_1c_2} \right) \psi_b = L^{ab}\psi_b,
\end{equation}
we observe that it such that $L^{ab}\psi_b = \gamma^{abc}\nabla^G_b\psi_c$, and so by applying a supersymmetry transformation
\begin{equation}
\label{eq:integrability cond2}
(\text{bosonic eoms})^a\cdot\varepsilon \propto \delta_{\varepsilon}(\text{fermionic eoms})^a = \delta_{\varepsilon} (L^{ab}\psi_b) = \delta_{\varepsilon}(\gamma^{abc}\nabla^G_b\psi_c) = \gamma^{abc}\nabla^G_b\nabla^G_c\varepsilon ,
\end{equation}
which implies~\eqref{eq:integrability cond}, as was remarked, for instance, in~\cite{Lukic:2007aj}. It was also noted in~\cite{Lukic:2007aj} that if the equations of motion are solved then $L \circ \nabla^G = 0$.  Therefore, when on-shell we have an exact sequence 
\begin{equation}
S \xrightarrow{\nabla^G} T^*\otimes S \xrightarrow{L} T^*\otimes S \xrightarrow{(\nabla^G)^{\dagger}} S .
\end{equation}
Here we write $S$ and $T^*$ for the spinor and cotangent bundles respectively and have used  $L=L^{\dagger}\Rightarrow (\nabla^G)^{\dagger}\circ L = 0$. As we just saw, the condition $L \circ \nabla^G = 0$ results from supersymmetry, while $L=L^{\dagger}$ can be derived from requiring the reality of $\int \bar{\psi}_a L^{ab} \psi_b$.

What appears to be less know is that if we write $(\tilde{\nabla}^G)^c = \tfrac19 \gamma_{a} L^{ac} = \gamma^{bc}\nabla^G_b $, i.e. the (left) gamma trace of the M-theory Rarita--Schwinger operator, then it is possible to recover the bosonic action from a  Lichnerowicz-type relation $(\tilde{\nabla}^G)^a \nabla^G_a\varepsilon \sim \mathcal{L}_B \varepsilon$. More exactly, from~\eqref{eq:integrability cond2} we see that $\tilde{\nabla}^G\nabla^G$ will be proportional to part of the bosonic  equations of motion $\gamma^{ab}\nabla^G_a\nabla^G_b\varepsilon \propto \text{(trace of Einstein equation}+\text{8-form gauge equation)}$ which can be used to reconstruct the bosonic action since $\mathcal{L}_B \sim (\text{trace of Einstein}) \vol_g +C\wedge\text{(8-form)}$, up to integration by parts.

Explicitly we have
\begin{equation}
\begin{aligned}
\label{eq:nablaF}
\nabla^G :\, &S \rightarrow T^*\otimes S,\\
&\nabla^G_a\varepsilon = \LC_a \varepsilon 
     + \tfrac{1}{288} \left(\gamma_a{}^{b_1\dots b_4}
        - 8 \delta_a{}^{b_1} \gamma^{b_2b_3b_4} 
        \right) G_{b_1\dots b_4} \varepsilon ,\\
\tilde{\nabla}^G :\, &T^*\otimes S \rightarrow S,\\
&\tilde{\nabla}^G_a \psi^a = \gamma^{ab}\nabla_a\psi_b +\tfrac{1}{144}\left(\gamma^{ab_1\dots b_4}-2\delta_a{}^{b_1} \gamma^{b_2b_3b_4}\right)G_{b_1\dots b_4} \psi_a ,
\end{aligned}
\end{equation}
and so $\tilde{\nabla}^G$ is the gamma-trace of gravitino equation of motion, i.e. it appears in the action in the term $\int \fsl{\bar{\psi}} (\tilde{\nabla}^G)^a	\psi_a$. Note also that $L=L^{\dagger} \Rightarrow (\tilde{\nabla}^G)^a\gamma_a =  \gamma_a((\tilde{\nabla}^G)^{\dagger})^a$. Then the Lichnerowicz relation is
\begin{equation}
\begin{aligned}
(\tilde{\nabla}^G)^a\nabla^G_a\varepsilon  &= -\tfrac14 \mathcal{R}\varepsilon + \tfrac{1}{24}\tfrac{1}{4!}G_{b_1\dots b_4} G^{b_1\dots b_4}\varepsilon -\tfrac{1}{144}(\nabla_{b_1}G_{b_2\dots b_5})\gamma^{b_1\dots b_5}\varepsilon\\
&\ph{:=}+\tfrac{1}{4}\tfrac13\tfrac{1}{3!}(\nabla^{b_1}G_{b_1\dots b_4})\gamma^{b_2b_3b_4}\varepsilon - \tfrac{1}{4}\tfrac16\tfrac{1}{4!^2}G_{b_1\dots b_4} G_{b_5\dots b_8}\gamma^{b_1\dots b_8}\varepsilon\\ 
&=-\tfrac14 \big(\mathcal{R} (\ast 1)_{b_1 \dots b_{11}} -\tfrac16 (G\wedge *G)_{b_1 \dots b_{11}} \big)\gamma^{b_1 \dots b_{11}}\varepsilon \\
&\ph{:=}-\tfrac{1}{4}\tfrac13\big(\tfrac{1}{7!}\nabla_{b_1}(\ast G)_{b_2\dots b_8}+ \tfrac12\tfrac{1}{4!^2}G_{b_1\dots b_4} G_{b_5\dots b_8}\big)\gamma^{b_1\dots b_8}\varepsilon ,
\end{aligned}
\end{equation}
where in the second equality we used the Bianchi identity for the flux. As expected, we obtained the trace of Einstein-Maxwell and the $C$ gauge equation of motion -- this relation is simply a result of applying a supersymmetry transformation to the  equation of motion for the trace of the gravitino.

Now the action
\begin{equation}
\begin{aligned}
\int\mathcal{L}_B &=\int \mathcal{R} \ast 1 -\tfrac{1}{2}G\wedge *G -\tfrac16 C\wedge G\wedge G ,\\
&=\int \mathcal{R}\ast 1 -\tfrac{1}{6}G\wedge *G -\tfrac13C\wedge \dd*G -\tfrac16 C\wedge G\wedge G + \text{boundary terms} ,\\
&=\int  \mathcal{R}\ast 1 -\tfrac{1}{6}G\wedge *G-\tfrac13C\wedge (\dd*G +\tfrac12 G\wedge G) + \text{boundary terms} ,
\end{aligned}
\end{equation}
so,  defining a polyform $\rho$ by $(\tilde{\nabla}^G\nabla^G\varepsilon) = -\tfrac14 \rho \cdot \varepsilon$, we have that the bosonic Lagrangian can be written compactly in terms of the Mukai pairing\footnote{The Mukai pairing is the top-form defined in $d$-dimensions from two polyforms $\alpha$ and $\beta$ by $<\alpha,\beta> = \sum_{p}  (\alpha^{(p)})^T\wedge \beta^{(d-p)}$, where $(\alpha^{(p)})^T_{a_1 \dots a_p}=\alpha^{(p)}_{a_p \dots a_1}$.} $\mathcal{L}_B = \rho|_{11} - C \wedge \rho|_{8} = < 1 +  C, \rho> 
$, up to integration by parts.

\subsection{Lichnerowicz Method}

This relation between the operator appearing in the supersymmetry variation of the gravitino and the bosonic action suggests we might be able to use it to build the bosonic sector of supersymmetric theories, short-cutting the usual Noether method. So let us invert the logic of the integrability calculations~\eqref{eq:integrability cond}, and consider a general type of differential operator acting on spinors. We will then promote this Lichnerowicz equation to a necessary constraint that the operator must satisfy, rather than have it be a consequence of supersymmetry.

Let us look at what this implies more concretely. We have a metric and a 4-form flux, and we want to build an operator that maps the supersymmetry parameter to the gravitino representation. Basic $SO(10,1)$ representation theory tells us there are two ways of tensoring a 4-form with a spinor and obtain a vector-spinor (essentially corresponding to embedding in the gamma-trace or gamma-traceless part of the vector-spinor), so we consider the following operators
\begin{equation}
\begin{aligned}
\Dgen :\, &S \rightarrow T^*\otimes S,\\
&\Dgen_a\varepsilon = \LC_a \varepsilon 
     + k_1 \gamma_a{}^{b_1\dots b_4}G_{b_1\dots b_4} \varepsilon 
      +k_2 \delta_a{}^{b_1} \gamma^{b_2\dots b_4} G_{b_1\dots b_4} \varepsilon ,\\
\tilde\Dgen :\, &T^*\otimes S \rightarrow  S,\\
&\tilde\Dgen^a \psi_a = \gamma^{ab}\nabla_a\psi_b + \tilde{k}_1\gamma^{ab_1\dots b_4}G_{b_1\dots b_4} \psi_a  + \tilde{k}_2g^{ab_1} \gamma^{b_2\dots b_4}G_{b_1\dots b_4} \psi_a,
\end{aligned}
\end{equation}
where $k_1$, $k_2$, $\tilde{k}_1$ and $\tilde{k}_2$ are left as arbitrary constants and note that a priori we may consider the `conjugate' operator that appears in the fermionic action to also be generic. 

Now we impose constraints to obtain a consistent Lichnerowicz type-relation from these operators. Clearly, if $\tilde\Dgen^a D_a\varepsilon$ is to define a tensor and not a differential operator, we must have that $\tilde{D}^b = \gamma^{ab}D_a$. This fixes $\tilde{k}_1$ and $\tilde{k}_2$ in terms of $k_1$ and $k_2$. 

We thus have that $\tilde{D}\circ D$ is a linear map $S \rightarrow S$, and so it must be a combination of $p$-forms. Then we have the physical requirement that this tensor should contain only scalars and 3-forms (up to Hodge dualisation, and after imposing the Bianchi identity for $G$), corresponding to the degrees of freedom of the trace of the metric and the 3-form gauge field (this is also clearly a necessary condition for $\tilde\Dgen \circ D = 0$  on-shell). This forces the ratio between $k_1$ and $k_2$ to be fixed, $k_2=-8k_1$. Therefore, from these simple constraints we are left with just one free coefficient, which can be absorbed in the normalisation of the fluxes. We recover $\Dgen = \nabla^G$ and as we have just seen the M-theory Lichnerowicz $\tilde\Dgen^a D_a\varepsilon$ will give the bosonic action.

\section{Higher-derivative terms}
\label{sec:high}

We will now try to apply this method to obtain the higher derivative effective action. A number of simplifications are assumed in what follows: we will discard terms which are higher order in derivatives than the order we are currently examining; we will ignore fluxes and Ricci terms, i.e. we will work only with the Riemann tensor which reduces to its Weyl tensor component; and we will also assume that  at the $R^4$ level the action does not contain `bare' connections, so we will not consider a term like $\nabla R\nabla R^2 $ for example.

Even with these simplifications, the form of the possible operators remains substantially less straightforward than in the classical flux case. In particular, in moving to higher order we should expect the corrections to include derivatives of the supersymmetry parameter or  the gravitino, cf.~\cite{Bergshoeff:1988nn,Peeters:2000qj}, which means that requiring that the M-theoretic Lichnerowicz equation defines a tensor is a bit more subtle. In particular, we will find we must abandon the classical relation  $\tilde{D}^a = \gamma^{ab} D_a$.

Consider, for example, a correction term corresponding to some tensor $ X_{abcd}$ which transforms in the representation $ X_{abcd} \in [0,2,0,0,0] $, that is, it has the same symmetry properties of the Weyl tensor. Note we are using the highest-weight notation of the computer algebra program LiE~\cite{LiE}, which we used extensively throughout this work for group theoretic calculations. We can define the following operators
\begin{equation}
\begin{aligned}
\Dgen :\, &S \rightarrow T^*\otimes S,\\
&\Dgen_a \varepsilon= \LC_a \varepsilon 
     + k_1 \left(\nabla^b X_{abcd}\right)\gamma^{cd} \varepsilon 
     + k_2  X_{abcd}\gamma^{cd} \nabla^b\varepsilon,\\
\tilde{D} :\, &T^*\otimes S \rightarrow S,\\
&\tilde{D}^a \psi_a  = \gamma^{ab}\left( \nabla_a\psi_b +\tilde k_1 \left(\nabla^c X_{acef}\right)\gamma^{ef} \psi_b
     + \tilde k_2 X_{acef}\gamma^{ef} \nabla^c \psi_b \right),
\end{aligned}
\end{equation}
where $k_1, k_2$ etc. parametrise higher-order corrections. 

Then, 
\begin{equation}
\begin{aligned}\label{eq:tensor}
(\tilde{D}\Dgen\varepsilon) &= \gamma^{ab}\nabla_a\nabla_b\varepsilon +\left(k_1-\tilde{k}_1-k_2\right)\left(\nabla^a X_{abcd}\right)\gamma^{cd}\nabla^b\varepsilon 
\\&\ph{:=}-k_1 \left(\nabla^a \nabla^b X_{abcd}\right)\gamma^{cd}\varepsilon 
-\left(\tilde k_2 + k_2\right) X_{abcd}\gamma^{cd}\nabla^a\nabla^b\varepsilon \\
&\ph{:=}+\text{higher order},
\end{aligned}
\end{equation}
so we have that, once we discard the higher order terms, tensoriality requires $k_1-\tilde k_1-k_2=0$. Note that it is crucial that the term with two covariant derivatives acting on the spinor is antisymmetrised on those $\nabla$, which is a consequence of the symmetry properties of $X_{abcd}$.\footnote{If the tensor $X_{abcd}$ were, say, fully symmetric, we would have extra constraints in order to ensure tensoriality. We would also not obtain extra Riemann tensors, instead we would be left with `naked' covariant derivatives.}

Solving the coefficient constraint for $\tilde{k}_1$, we are left with
\begin{equation}
\begin{aligned}
(\tilde{D}\Dgen\varepsilon) &=-\tfrac14 \mathcal{R}\varepsilon+\tfrac12\left(2k_1-\tilde k_2 - k_2\right) R^{abe}{}_c X_{abed}\gamma^{cd}\varepsilon \\
&\ph{:=}+\tfrac14\left(\tilde k_2 + k_2\right)R^{abcd} X_{abcd}\varepsilon
-\tfrac18\left(\tilde k_2 + k_2\right) R^{ab}{}_{cd}X_{abef}\gamma^{cdef}\varepsilon  \\
&\ph{:=}+\text{higher order} .
\end{aligned}
\end{equation}
So, whereas we would like $\tilde{D}$ to be determined by $D$, we are left with ambiguities. 

The choice we will be making from now on is to always take $\tilde k_1=0$, $k_1=k_2=\tilde k_2$. This means that both $\tilde{D}^a\nabla_a\varepsilon$ and $\gamma^{ab}\nabla_aD_b\varepsilon$ are separately tensorial, which can be interpreted as writing the fermionic action in terms of ``supercovariant" objects~\cite{Peeters:2000qj}. We thus have that $\tilde{D}$ is completely fixed given a $D$ to a certain order, and so, writing explicit spinor indices, we will consider operators in the form
\begin{equation}
\begin{aligned}
 D_a \varepsilon^{\alpha} &= \nabla_a \varepsilon^{\alpha} + k \nabla_b \big( \Theta_{a}{}^b{}_{\beta}{}^{\alpha}\varepsilon^{\beta}\big) = \nabla_b \Big(  \big(  \mathbb{I}_{a}{}^b{}_{\beta}{}^{\alpha}  + k \Theta_{a}{}^b{}_{\beta}{}^{\alpha}\big)\varepsilon^{\beta}\Big) ,\\
 \tilde{D}^a\psi_a{}^{\alpha} &=  \gamma^{ac}{}_{\beta}{}^{\alpha}\big(\nabla_a \psi_c
 + k   \Theta_{a}{}^b{}_{\gamma}{}^{\beta} \nabla_b\psi_c{}^{\gamma}\big) =  \gamma^{ac}{}_{\beta}{}^{\alpha}\big(\mathbb{I}_{a}{}^b{}_{\gamma}{}^{\beta} 
 + k   \Theta_{a}{}^b{}_{\gamma}{}^{\beta} \big)\nabla_b\psi_c{}^{\gamma},
 \end{aligned}
\end{equation}
with $\Theta$ some object that is a function of (powers of) the Riemann curvature. 

\subsection{$R^2$ and $R^3$ couplings in the effective action}
\label{sec:R2+R3}

It is well known that the first corrections to the eleven-dimensional action begin at $R^4$. However, as a simple exercise let us check whether we can define operators that through Lichnerowicz would be compatible with an $R^2$ or $R^3$ action.

Counting derivatives, we see that in order to obtain an $R^2$ correction to the action, there is only one type of term we can add to the operators, namely
\begin{equation}
\begin{aligned}  \label{eq:r2susy}
&\Dgen_a\varepsilon = \LC_a \varepsilon 
     + k \nabla^b \left(R_{abcd}\gamma^{cd} \varepsilon\right) ,\\
&\tilde{D}^a \psi_a = \gamma^{ab}\nabla_a\psi_b 
     + k  R_{abcd}\gamma^{cd} \nabla^b \psi_a .
\end{aligned}
\end{equation}
However, it is clear that $\tilde{D}^aD_a\varepsilon$ will have a 4-form, $k R_{a_1 a_2 b_1 b_2}R_{a_3 a_4 b_1 b_2}\gamma^{a_1 a_2 a_3 a_4} \varepsilon$, so satisfying the constraint that $\tilde{D}D\varepsilon$ should contain only scalars and 3-forms forces $k=0$. We conclude no corrections are admissible at this order.

For $R^3$ there are more possibilities. By derivative counting we have that the extra pieces in $D$ must be of the type $\nabla R^2$. So we look at the tensor decomposition of $\otimes^2 R$ and find that the families of terms that can be considered in the operators are the ones listed in table~\ref{tab:R3}.

\begin{table}[htb]
\begin{center}
\begin{tabular}{ccc}
Projection of $R^2$& Rep  of $SO(10,1)$ & Multiplicity \\
\hline
\hline
$\hat{X}^i$ & [0,2,0,0,0] & 2 \\
$\hat{W}$ & [2,0,0,0,0] & 1 \\
$\hat{S}$ & [0,0,0,0,0] & 1 \\ \hline
$\hat{T}$ & [0,0,0,1,0] & 1 
\end{tabular}
\end{center}
\caption{Valid embeddings of $\otimes^2 R$ in $\delta\psi$.} 
\label{tab:R3}
\end{table}

Explicitly, these decompositions are given by
\begin{equation}
\begin{aligned}
\hat{X}^1_{a_1 a_2 c_1 c_2} &=\tfrac12 R_{a_1 a_2 b_1 b_2} R_{c_1 c_2 b_1 b_2} + \tfrac12 R_{a_1 c_2 b_1 b_2} R_{c_1 a_2 b_1 b_2} \\&\phantom{:=} -\tfrac13 g_{a2 c2} R_{a_1 d_1 b_1 b_2} R_{ c_1 d_1 b_1 b_2} +\tfrac{1}{60} g_{a1 c1}g_{a2 c2} R_{d_1 d_2 b_1 b_2} R_{ d_1 d_2 b_1 b_2}  ,\\
\hat{X}^2_{a_1 a_2 c_1 c_2} &= \tfrac12 R_{a_1  b_1 c_1 b_2} R_{a_2 b_1 c_2 b_2} + \tfrac12 R_{c_1  b_1 a_1 b_2} R_{a_2 b_1 c_2 b_2} \\&\phantom{:=} +	\tfrac16 g_{a2 c2} R_{a_1 d_1 b_1 b_2} R_{ c_1 d_1 b_1 b_2} - \tfrac{1}{120} g_{a1 c1}g_{a2 c2} R_{d_1 d_2 b_1 b_2} R_{ d_1 d_2 b_1 b_2} ,\\
\hat{W}_{a_1 c_1} &= R_{a_1 d_1 b_1 b_2} R_{ c_1 d_1 b_1 b_2}-\tfrac{1}{11}g_{a_1 c_1} R_{d_1 a_2 b_1 b_2} R_{d_1 a_2 b_1 b_2} ,\\
\hat{S} &= R_{a_1 a_2 b_1 b_2} R_{a_1 a_2 b_1 b_2},\\
\hat{T}_{a_1 a_2 a_3 a_4} &= R_{a_1 a_2 b_1 b_2} R_{a_3 a_4 b_1 b_2} ,
\end{aligned}
\end{equation}
where same-letter free indices are assumed to be antisymmetrised. 
However, the $\hat{W}$ and $\hat{S}$ do not actually contribute to the Lichnerowicz (they are projected out) so we will not consider them as they cannot possibly source corrections to the action.

There is now a new potential ambiguity in how to construct the supersymmetry operators, as there are several different ways of embedding the $\hat{T}$ term in $\delta\psi$. We will discuss this in more detail in the $R^4$ section, but some of these can be fixed by consistency conditions from the Lichnerowicz equation, and the rest by requiring that the $\hat{T}$ embed strictly into the traceless-vector-spinor in the conjugate operator $\tilde{D}$, with the trace taken from the right so that the operator $\tilde{\slashed{D}}$ is now the one that appears in the fermionic coupling of the gravitino trace $\int \fsl{\bar{\psi}}\tilde{\slashed{D}}\fsl{\psi}$. This will imply that $\tilde{D}^a \gamma_a = \nabla^a\gamma_a$, as the $\hat{X}^i$ likewise only embed in the traceless part. This also ensures that $\tilde{\slashed{D}}=\tilde{\slashed{D}}^{\dagger} $.

We end up with
\begin{equation}\label{eq:DR3}
\begin{aligned}
\Dgen_a\varepsilon &= \LC_a \varepsilon + \sum_{i=1}^2 \hat{x}_i \gamma^{cd}  \nabla^b ( \hat{X}^i_{abcd} \varepsilon )\\
       &\ph{:=}+    \hat{t}\,\Big(  
      \gamma^{b_1b_2}\nabla^{b_3}(\hat{T}_{ab_1 b_2 b_3}\varepsilon  ) 
      +    \tfrac{1}{14}\gamma^{b_1\dots b_4} \nabla_a (\hat{T}_{b_1\dots b_4}\varepsilon)    
        +    \tfrac{1}{14}\gamma^{b_1\dots b_4}\nabla_{b_1} (\hat{T}_{ab_2\dots b_4}\varepsilon) \\    
      &\ph{:=} +\tfrac{1}{7} \gamma_a{}^{b_2b_3 b_4}\nabla^{b_1} (\hat{T}_{b_1\dots b_4}\varepsilon)
      +\tfrac{1}{84}\gamma_a{}^{b_1\dots b_5}\nabla_{b_5} (\hat{T}_{b_1\dots b_4}\varepsilon) 
      \Big),
\end{aligned}
\end{equation}
and
\begin{equation}
\begin{aligned}
\tilde{D}^a\psi_a&= \gamma^{ab} \LC_a \psi_b  + \sum_{i=1}^2 \hat{x}_i \gamma^{cd}   \hat{X}^i_{abcd}   \nabla^b\psi^a
       \\&\ph{:=}+  \hat{t}\, \Big(\tfrac52 \gamma^{c_1c_2} \hat{T}_{abc_1c_2}  \nabla^{a}\psi^b  
       +\tfrac{5}{14} \gamma^{a c_2 c_3 c_4}\hat{T}_{c_1\dots c_4}\big( \nabla_a\psi^{c_1} -  \nabla^{c_1}\psi_a \big) 
     \\&\ph{:=} -\tfrac{5}{84} \gamma^{abc_1\dots c_4}\hat{T}_{c_1\dots c_4} \nabla_a\psi_b 
      \Big),
\end{aligned}
\end{equation}
so we have a priori three free coefficients, $\hat{x}_1,\hat{x}_2,\hat{t}$. If we compute $\tilde{D}^aD_a\varepsilon$ we find that in addition to scalars in $R^3$, the Lichnerowicz also contains 4-forms, which must be cancelled. There are exactly three possible independent $R^3$ 4-forms, all of which appear in computation, and eliminating  them forces precisely that all $\hat{x}_i= \hat{t}=0$. We conclude that no $R^3$ corrections are admissible.

\subsection{$R^4$ couplings}

Finally, we arrive at corrections to the supersymmetry operators that may lead to $R^4$ corrections to the action via the Lichnerowicz procedure. We will need to add $\nabla R^3$ terms to $\delta \psi$, for which there are several possibilities depending on how we build the $R^3$ factor, as listed in table~\ref{tab:R4a}.
\begin{table}[htb]
\begin{center}
\begin{tabular}{ccc}
\\
Projection of $R^3$ & Rep  of $SO(10,1)$ & Multiplicity \\
\hline
\hline
$X^i$ & [0,2,0,0,0] & 8 \\
$W^i$ & [2,0,0,0,0] & 3 \\
$S^i$ & [0,0,0,0,0] & 2 \\ \hline
$Y^i$ & [0,1,0,0,2] & 2 \\
$V^i$ & [1,0,0,0,2] & 2 \\ 
$T^i$ & [0,0,0,1,0] & 3 \\ \hline
$Z^i$ & [0,1,0,1,0] & 3 \\
$U^i$ & [1,0,1,0,0] & 3 \\ \hline
$L^i$ & [2,1,0,0,0] & 3 \\
$M^i$ & [2,0,0,1,0] & 6 
\end{tabular} \\
\end{center}
\caption{Valid embeddings of $\otimes^3 R$ in $\delta\psi$. In each case, the index $i$ runs over the corresponding multiplicity.} 
\label{tab:R4a}
\end{table}

We find these families of terms  by decomposing into irreducible representations   the product of three Weyl tensors, i.e. if we denote the Weyl curvature representation $[0,2,0,0,0]$ by $W$, we consider the symmetrised cubic tensor product $S^3 W$. Then we restrict to those which, when combined with a $\nabla_a$ and a spinor $\varepsilon$, can contain a vector spinor $\delta\psi_a$. In other words, we look for the irreducible representations in the intersection $S^3 W \cap (V\otimes S)^* \otimes (V\otimes S)$. We then verify these projections by explicit construction -- they are given in appendix~\ref{app:projR4}.

Note that in some cases there are multiple ways of embedding these terms into a vector spinor. For example, consider the three  $T^i$, each of which is a 4-form. There are five inequivalent ways of combining a 4-form together with $\nabla_a$ and $ \varepsilon$ to obtain a valid $\delta \psi_a$. These roughly correspond to using either a $\gamma^{(2)}$, a  $\gamma^{(4)}$ or a  $\gamma^{(6)}$ to ``soak up'' the indices, and then further distinguishing in the first two whether the overall free index is symmetrised or antisymmetrised with the index on $\nabla_a$. So a priori, the $T^i$ would appear to contribute 15 undetermined coefficients to our operators.

There are, however, some constraints from the M-theoretic Lichnerowicz that can be immediately applied to reduce the number of possibilities. We already established we only want antisymmetrised products of $\nabla$ appearing in $\tilde{D}^a D_a \varepsilon$, so that they will be converted into a Riemann tensor. Symmetrised $\nabla$ would lead to `bare' connections in the action and could even spoil the tensoriality of the Lichnerowicz. This constraint eliminates the last two terms in the table, $L^i$ and $M^i$, and further reduces the admissible embeddings of some of the other terms. For instance, the $T^i$ can now each only be embedded in three different ways, for a total of 9 free coefficients. 

Finally, we impose the condition that only the traceless part of the $\tilde{D}$ operator gets modified which, as mentioned in the previous subsection, is sufficient to ensure that $\tilde{\slashed{D}}=\tilde{\slashed{D}}^{\dagger} $. It turns out that this is enough to fix these ambiguities of embedding, and the total number of remaining free coefficients matches precisely the multiplicity of projections of $R^3$.

Now, the representation theory also tells us what type of $p$-forms to expect in the Lichnerowicz. Since we have ensured that this expression will be, schematically, $  R^3 [\nabla,\nabla] \varepsilon = R^4 \varepsilon$, we need to look at the $p$-forms in the symmetric tensor product of four Weyl curvatures and see how each of the remaining terms in table might contribute to them. We summarise the result in table~\ref{tab:R4b}.

\begin{table}[htb]
\begin{center}
\begin{tabular}{c|cccccc}
$p$-form :	& 0 & 1 & 2 & 3 & 4 & 5\\
\hline
\hline
 $R^4$ multiplicity:	& 7 & 0 & 1 & 2 & 17 & 0\\
\hline
\hline
$X^i\otimes R$ & $\bullet$ & - &  $\bullet$ & - &  $\bullet$ & -\\
$W^i\otimes R$ & - & - & - & - & - & -\\
$S^i\otimes R$ & - & - & - & - & - & -\\ \hline
$Y^i\otimes R$ & - & - & - &  $\bullet$ &  $\bullet$ & -\\
$V^i\otimes R$ & - & - & - & - &  $\bullet$ & -\\ 
$T^i\otimes R$ & - & - & - & - &  $\bullet$ & -\\ \hline
$Z^i\otimes R$ & - & - &  $\bullet$ & - &  $\bullet$ & -\\
$U^i\otimes R$ & - & - &  $\bullet$ & - &  $\bullet$ & -\\ 
\end{tabular} \\
\end{center}
\caption{Potential contributions of the terms in table~\ref{tab:R4a} to different $p$-forms in the M-theoretic Lichnerowicz.} 
\label{tab:R4b}
\end{table}

We immediately observe that, as previously remarked, terms of type $W^i$ and $S^i$ are projected out in the Lichnerowicz computation, and so their presence (or lack thereof)  in the supersymmetry operator cannot be constrained in this manner. We also see that we need to have $X^i$ type terms if we hope to find corrections to the scalars, and $Y^i$ for corrections to the 3-form. These are the only objects that may contribute to a corrected action. However, they will also give rise to 4-forms which cannot be cancelled solely by picking appropriate coefficients $x_i$ and $y_i$ for each of them. Instead, we find by explicit computation that we must, at a minimum, also add terms of type $V^i$  and $T^i$, as we will now describe. 

\subsubsection{Minimal solution}
\label{sec:11d-r4-min}

Given these ingredients, we build:
\begin{equation}
\begin{aligned}
\label{eq:Dmin}
\Dgen :\, &S \rightarrow T^*\otimes S,\\
\Dgen_a\varepsilon &= \LC_a \varepsilon + \sum_{i=1}^8 x_i \gamma^{cd}  \nabla^b ( X^i_{abcd} \varepsilon )
       + \sum_{i=1}^2 y_i\gamma^{c_1\dots c_6}   \nabla^b( Y^i_{abc_1\dots c_6} \varepsilon )\\
       &\ph{:=}+ \sum_{i=1}^3   t_i\Big( 
      \gamma^{b_1b_2}\nabla^{b_3}(T^i_{ab_1 b_2 b_3}\varepsilon  ) 
      +    \tfrac{1}{14}\gamma^{b_1\dots b_4} \nabla_a (T^i_{b_1\dots b_4}\varepsilon)    
        +    \tfrac{1}{14}\gamma^{b_1\dots b_4}\nabla_{b_1} (T^i_{ab_2\dots b_4}\varepsilon) \\    
      &\ph{:=}+\tfrac{1}{7} \gamma_a{}^{b_2b_3 b_4}\nabla^{b_1} (T^i_{b_1\dots b_4}\varepsilon)
      +\tfrac{1}{84}\gamma_a{}^{b_1\dots b_5}\nabla_{b_5} (T^i_{b_1\dots b_4}\varepsilon)    
      \Big)\\
       &\ph{:=}+ \sum_{i=1}^2   v_i\Big(  \tfrac{1}{4}\gamma^{b_1\dots b_4}\nabla^{b_5}( V^i_{ab_1\dots b_5}\varepsilon - V^i_{b_5ab_1\dots b_4} \varepsilon) +\tfrac{1}{25}\gamma_a{}^{b_1\dots b_5}\nabla^{c}( V^i_{cb_1\dots b_5}\varepsilon)
      \\&\ph{:=}+ \tfrac{1}{25}\gamma^{b_1\dots b_6} \nabla_{b_6}(V^i_{ab_1\dots b_5}  \varepsilon)            
      \Big).
\end{aligned}
\end{equation}
As mentioned, these combinations ensure that Lichnerowicz will not contain symmetrised $\nabla$. In other words, we have that $\tilde{D}D\varepsilon = \sum_n (R^4)_{a_1\dots a_n}\gamma^{a_1\dots a_n}\varepsilon$ with no `bare' connections left. This should be clear once we look at the conjugate $\tilde{D}$ operator:
\begin{equation}
\begin{aligned}
\label{eq:tDmin}
\tilde{D} :\, &T^*\otimes S \rightarrow S,\\
\tilde{D}^a\psi_a&= \gamma^{ab} \LC_a \psi_b  + \sum_{i=1}^8 x_i \gamma^{cd}   X^i_{abcd}   \nabla^b\psi^a
       + \sum_{i=1}^2 y_i \gamma^{c_1\dots c_6}Y^i_{abc_1\dots c_6}   \nabla^b \psi^a 
       \\
       &\ph{:=}+ \sum_{i=1}^3  t_i \Big(\tfrac52 \gamma^{c_1c_2} T^i_{abc_1c_2}  \nabla^{a}\psi^b  
       +\tfrac{5}{14} \gamma^{a c_2 c_3 c_4}T^i_{c_1\dots c_4}\big( \nabla_a\psi^{c_1} -  \nabla^{c_1}\psi_a \big) 
      -\tfrac{5}{84} \gamma^{abc_1\dots c_4}T^i_{c_1\dots c_4} \nabla_a\psi_b 
      \Big)
       \\
       &\ph{:=}+ \sum_{i=1}^2    v_i \Big(  \gamma^{ac_3 c_4 c_5}V^i_{ac_1\dots c_5} \nabla^{c_1}\psi^{c_2}
       +\tfrac{1}{25}\gamma^{bc_1\dots c_5}V^i_{ac_1\dots c_5} \big(  \nabla_b \psi^{a} -    \nabla^a\psi_{b} \big)      
      \Big).
\end{aligned}
\end{equation}
Note that $\tilde{D}^a\gamma_a = \slashed{\nabla}$ as required.

Now we need to fix the coefficients such that only scalars and 3-forms survive. We used the computer program Cadabra~\cite{Cadabra} to solve this algebra problem (and also performed some double-checks on Mathematica). The single possible $R^4$ 2-form  vanishes identically and eliminating the 4-forms is solved by:
\begin{equation}
\begin{aligned}
	& x_1 = - y_2 ,	\quad 	& x_7 &= 16 y_2 ,\quad 	& t_2 &= \tfrac{2}{45} ( 72 y_2 - x_5  ), \\ 
 	& x_2 = 16 y_2 ,	\quad 	& x_8 &= 16 y_2 ,\quad  & t_3 &= -\tfrac{4}{45} ( 72 y_2 - x_5  ),\\
	& x_3 = -2  y_2 ,	\quad  	& y_1 &= -\tfrac14 y_2 ,\quad  & v_2 &= 	\tfrac{5}{8} ( -24 y_2 + 5 x_5  ) ,\\
 	& x_4 = -16 y_2 , 	\quad  	& t_1 &= 0, \\
	& x_6 = -4 x_5 , 	\quad  	& v_1 &= 0, \\
\end{aligned}
\end{equation}
with $y_2,x_5$ free. We have therefore found two solutions for our operators at $R^4$ such that the Lichnerowicz contains only scalars and 3-forms (or rather, what results naturally from the computation are their Hodge-dual 8-forms)
\begin{equation}
\tilde{D}^aD_a \varepsilon = (\text{scalar})\varepsilon + (\text{8-form})_{a_1 \dots a_8}\gamma^{a_1 \dots a_8}\varepsilon .
\end{equation}

Note that in the notation of~\cite{Peeters:2000qj, Antoniadis:2003sw}, we have a basis for the seven different $R^4$ scalars
\begin{equation}\label{eq:R4-scalars}
\begin{aligned}
R\cdot X^1 & = R_{44},\quad & R\cdot X^2 & = R_{45},\quad& R\cdot X^3 & = R_{43},\\
R\cdot X^4 & = R_{41},\quad& R\cdot X^5 & = R_{46},\quad& R\cdot X^6 & = A_7,\\
R\cdot X^7 &  = - R_{42} -\tfrac14 R_{46} + A_7,\quad & R\cdot X^8 & =  - R_{42} +\tfrac14 R_{46} + A_7 ,
\end{aligned}
\end{equation}
where $R\cdot X^i = R^{abcd}X^i_{abcd}$ and so the scalar component of the M-theory Lichnerowicz can be written as (below we write the Hilbert-Einstein term for convenience;  other terms containing the Ricci tensor or scalar  are dropped)
\begin{equation}
\begin{aligned}
\label{eq:LL11D}
(\tilde{D}^aD_a\varepsilon|_{\text{scalar}})&=-\tfrac14 \mathcal{R}
+\tfrac12 x_5(  R_{46} -4   A_7)\\
 &\ph{:=}+\tfrac{1}{2}  y_2( -16  R_{41} -32 R_{42} -2 R_{43} -  R_{44} +16 R_{45} +32 A_7) \\
&= -\tfrac14 \mathcal{R}
-\tfrac12 \tfrac{1}{192 }x_5(  -192 R_{46} +768   A_7)\\
 &\ph{:=}-\tfrac{1}{2} \tfrac{1}{12} y_2( 192  R_{41} +384 R_{42} +24 R_{43} +12  R_{44}-192 R_{45} - 384 A_7) \\
&= -\tfrac14 \mathcal{R} \\&\ph{:=}
-\tfrac14 \tfrac{1}{96 }x_5(t_8 t_8 R^4 -\tfrac14 E_8 )
 -\tfrac{1}{4} \tfrac{1}{12} y_2(t_8 t_8  R^4 +\tfrac14 E_8 ) \ ,
\end{aligned}
\end{equation}
while the 8-form is:
\begin{equation}
\begin{aligned}
(\tilde{D}^aD_a\varepsilon|_{\text{8-form}})_{a_1 \dots a_8} &=- \tfrac{1}{4} y_2(  -\tfrac14 R\cdot Y^1 +  R\cdot  Y^2)_{a_1 \dots a_8} \\
& = - \tfrac{1}{4} y_2( - \tfrac{1}{4} R_{a_1 a_2 b_1 b_2} R_{a_3 a_4 b_1 b_2} R_{a_5 a_6 c_1 c_2} R_{a_7 a_8 c_1 c_2} \\&\ph{= - \tfrac{1}{4} y_2( }+  R_{a_1 a_2 b_1 c_1 } R_{a_3 a_4 c_1 c_2} R_{a_5 a_6 c_2 b_2  }  R_{a_7 a_8 b_2 b_1 }),
\end{aligned}
\end{equation}
so $\tfrac{1}{16}  R\cdot Y^1  = \tr R^2 \wedge \tr R^2$ and $ \tfrac{1}{16} R\cdot Y^2  = \tr R^4 $. We thus recognise the $y_2\neq 0$ solution as corresponding to the invariant $(t_8 t_8R^4 +\tfrac14 E_8 )$ and $x_5\neq 0$ to $(t_8 t_8R^4 - \tfrac14 E_8 )$. To get the correct eleven-dimensional action we thus need to take $x_5=0$ and $y_2$ will be fixed by the normalisation of the higher derivative terms. The existence of this freedom of choice is maybe not unexpected, as three different superinvariants had already been identified previously~\cite{deRoo:1992zp}. Comparing with~\cite{Peeters:2000qj} we have, in the notation of that paper, that the $y_2$ terms give the $I_X+\tfrac18 I_Z$ invariant, while the $x_5$ terms give $I_X-\tfrac18 I_Z$.

In appendix~\ref{app:non-min} we discuss other possible solutions to this Lichnerowicz system.

\section{Application: seven-dimensional internal spaces}
\label{sec:7D}

Two out of the three higher-derivative structures allowed by supersymmetry involve a complete antisymmetrisation of eight indices. Hence, for our computation, seven is a sort of critical dimension, i.e. the highest where only a single structure survives. So, in addition to the natural desire of learning more about M-theory compactifications on seven-dimensional internal spaces, assuming the eleven-dimensional space breaks as $\mathcal{M}_{11}=\mathcal{M}_{4}\times \mathcal{M}_{7}$ and focusing on the $\mathcal{M}_7$ component provides a good setting to make further considerations on our construction.\footnote{Of course, a calculation on $\mathcal{M}_7$ can be considered on its own, with the view of deriving higher-derivative corrections in seven-dimensional effective theories. We mostly concentrate on supersymmetry on the internal seven-manifold.}
We then have that the two eleven-dimensional scalars $y_2$ and $x_5$ will coincide $(t_8 t_8  R^4 + \tfrac14 E_8)  =(t_8 t_8 R^4 - \tfrac14 E_8)  = t_8 t_8 R^4 $ when fully restricted to $\mathcal{M}_7$, and even though the $y_2$ is the correct physical solution in eleven dimensions, in seven dimensions we may just as well work with the simpler $x_5$ when writing a fully restricted operator $D$.  In terms of $\SO(7)$ representations using the projectors of the cubic powers of $R_{mnpq}$ in appendix~\ref{app:projR4-7}, this is

\begin{equation}
\begin{aligned}
\label{eq:X5reduced}
\Dgen_m\varepsilon &= \nabla_m \varepsilon +  x_5 \gamma^{pq}  \nabla^n \Big( ( \check X^5_{mnpq} -4   \check X^6_{mnpq} ) \varepsilon \Big)+\tfrac{2}{315} x_5 \gamma_m{}^n\nabla_n\Big( (\check{S}^1
-2   \check{S}^2) \varepsilon\Big)\\
&\phantom{:=}+\tfrac{8}{45}x_5  \gamma^{np} \nabla_n  \Big( ( -\check{W}^1_{mp}
+2 \check{W}^2_{mp}  
 + \check{W}^3_{mp} ) \varepsilon\Big)   -\tfrac{8}{45} x_5 \gamma_m{}^n\nabla^p\Big( (-\check{W}^1_{pn}
+2 \check{W}^2_{pn} + \check{W}^3_{pn}) \varepsilon\Big) \\
       &\ph{:=}+ x_5 \Big( 
      - \tfrac{2}{45}\gamma^{n_1n_2}\nabla^{n_3}(\check T^2_{mn_1 n_2 n_3}\varepsilon  ) 
      - \tfrac{1}{315}\gamma^{n_1\dots n_4} \nabla_m (\check T^2_{n_1\dots n_4}\varepsilon)    
        -\tfrac{379}{1260} \gamma^{n_1\dots n_4}\nabla_{n_1} (\check T^2_{mn_2\dots n_4}\varepsilon ) \\     
      &\ph{:=}+\tfrac{367}{1260} \gamma_m{}^{n_2n_3 n_4}\nabla^{n_1} (\check T^2_{n_1\dots n_4}\varepsilon)
      -\tfrac{113}{945}  \gamma_m{}^{n_1\dots n_5}\nabla_{n_5} (\check T^2_{n_1\dots n_4}\varepsilon)    
      \Big) \\
        &\ph{:=}+  x_5 \Big( 
      \tfrac{4}{45}\gamma^{n_1n_2}\nabla^{n_3}(\check T^3_{mn_1 n_2 n_3}\varepsilon  ) 
      +   \tfrac{2}{315} \gamma^{n_1\dots n_4} \nabla_m (\check T^3_{n_1\dots n_4}\varepsilon)    
       +\tfrac{11}{45} \gamma^{n_1\dots n_4}\nabla_{n_1} (\check T^3_{mn_2\dots n_4}\varepsilon) \\       
      &\ph{:=}-\tfrac{71}{315}  \gamma_m{}^{n_2n_3 n_4}\nabla^{n_1} (\check T^3_{n_1\dots n_4}\varepsilon)
      +\tfrac{13}{135}\gamma_m{}^{n_1\dots n_5}\nabla_{n_5} (\check T^3_{n_1\dots n_4}\varepsilon)    
      \Big) \\
       &\ph{:=}+\tfrac{25}{8} x_5 \Big(  \tfrac{1}{4}\gamma^{n_1\dots n_4}\nabla^{n_5}( \check{V}^2_{m  n_1 n_2 n_3 n_4 n_5} \varepsilon - \check{V}^2_{ n_5 m  n_1 n_2 n_3 n_4}  \varepsilon) 
  \\&\ph{:=}+\tfrac{1}{25}\gamma_m{}^{n_1\dots n_5}\nabla^{p}(  \check{V}^2_{p  n_1 n_2 n_3 n_4 n_5} \varepsilon) + \tfrac{1}{25}\gamma^{n_1\dots n_6} \nabla_{n_6}( \check{V}^2_{m  n_1 n_2 n_3 n_4 n_5} \varepsilon)           
      \Big) ,
\end{aligned}
\end{equation}
where $m,n,\dots$ are internal seven-dimensional indices and now $\nabla$ is the Levi-Civita of the internal manifold and $R_{mnpq}$ its (Weyl) curvature.

The full decomposition, keeping dependence on both internal and external contributions, will be much more involved and we will not perform it here. However, we can try to capture some aspects by considering a restricted Lichnerowicz formula in the seven-dimensional space, expanded in order of \emph{internal} derivatives. As we will soon see, this seven-dimensional $\LL$ formula has corrections starting from three-derivatives on the operators and will turn out to reveal some  interesting new structures. In order to relate it to its eleven-dimensional counterpart, one needs to think of the coefficients in these operators as being made of Riemann curvatures in the external (four-dimensional) spacetime.\footnote{In order to make proper contact with the eleven-dimensional calculation, one should strictly start with a zero-derivative term $D_m\varepsilon=\nabla_m\varepsilon+\Lambda\gamma_m\varepsilon$, corresponding to taking all the $R^4$ couplings in the external four-dimensional space, and which would lead to a constant piece in the internal action $\tilde{D}^mD_m\varepsilon=-\tfrac14(\mathcal{R}+168\Lambda^2)\varepsilon$. To keep the calculation similar to the eleven-dimensional one, we will ignore this term.}

\subsection{M-theory Lichnerowicz for $\SO(7)$}
\label{ssec:ML}

In performing the M-theory Lichnerowicz in seven dimensions, we may immediately expect one difference. The eleven-dimensional $\LL$ was picking zero- and eight-forms in the expansion of $\rho$. In seven-dimensions, one is instead interested in zero- and four-form parts and the action, up to integration by parts, is given by:
 \begin{equation}
 \label{eq:7daction}
 \mathcal{L}_B = \rho|_{7} - C \wedge \rho|_{4} = < 1 +  C, \rho>   .
 \end{equation}

\subsubsection{Three-derivative terms in the operators}

In section~\ref{sec:R2+R3} we used a very simple argument to rule out any four-derivative contributions to the eleven-dimensional action -- they would inevitably contribute an unphysical 4-form to the M-theory Lichnerowicz. However, in seven dimensions a 4-form is Hodge-dual to a 3-form and thus contributes to the flux equation of motion. Therefore, we will allow the operators 
\begin{equation}
\label{eq:3der}
\begin{aligned} 
&\Dgen_m\varepsilon = \LC_m \varepsilon 
     + k \nabla^n \left(R_{mnpq}\gamma^{pq} \varepsilon\right) ,\\
&\tilde{D}^m \psi_m = \gamma^{mn}\nabla_m\psi_n 
     + k  R_{mnpq}\gamma^{pq} \nabla^n \psi_m .
\end{aligned}
\end{equation}
 These operators are not in contradiction with the eleven-dimensional construction -- in fact, they should be expected. Consider the $X^1$ term in the solution~\eqref{eq:Dmin}. Explicitly $\nabla^b X^1_{abcd} = \nabla^b R^{\tiny \mbox{11d} }_{abcd} (R^{\tiny \mbox{11d} })^2$ and one way of decomposing this is as $\nabla^n R^{\tiny \mbox{7d} }_{mnpq} (R^{\tiny \mbox{4d} })^2 \mapsto k \nabla^n R^{\tiny \mbox{7d} }_{mnpq}   $. 
 So consistency with our eleven-dimensional solution implies that $k$ is of at least quartic order in \emph{external} derivatives. 
 
The seven-dimensional Lichnerowicz now yields
\begin{equation}
\begin{aligned}
\label{eq:LL4der}
\tilde{D}^mD_m\varepsilon = -\tfrac14 \mathcal{R}\varepsilon +\tfrac12 k R_{mnpq}R^{mnpq}\varepsilon -\tfrac14 k R_{mnpq}R_{rs}{}^{pq}\gamma^{mnrs} \varepsilon + \text{ higher order}.
\end{aligned}
\end{equation}
In the effective theory the interpretation is immediate: using \eqref{eq:7daction} we obtain the $C_3 \wedge (\tr R^2)$ term of the theory with 16 supercharges \cite{DLM} together with its (Riemann)$^2$ completion.
As for the internal supersymmetry, we may already note a major limitation of our approach. Since we are effectively integrating out the four-dimensional action, we lost the ability to distinguish a four-dimensional scalar from a top-form. Indeed it is not hard to see that $X^1_{abcd} $ is not the only source of term $\sim k$ in \eqref{eq:3der}.  Such a term may also originate from $Y^1_{a_1a_2b_1\dots b_6}$. From other side, one may already guess that like in the seven-dimensional effective theory, these will yield internal top-forms quartic in derivatives. We shall return to these in subsection \ref{ssec:dis}.

Before turning to the discussion of contributions with a higher number of derivatives, we should remark that 
this this interpretation results in an extra physical constraint, which is important for reducing the number of  further terms. As mentioned in section~\ref{sec:11d-rev}, $X_8$ is related to M5/NS5 anomalies and string-theoretically it is one-loop and does not receive any higher loop contributions. On the other hand,  the next set of corrections to the internal covariant derivatives are, from the eleven-dimensional point of view, reductions of terms which are (at least) thirteen derivatives, i.e. the types of corrections that may contribute to $R^7$ (and higher) couplings. These come from string two (and higher) loops, and hence should not affect $X_8$.   This means that there can be no further higher-derivative 4-forms in seven dimensions, as these would correct $X_8$ when lifted to eleven dimensions. So even if there exist six- (or higher) derivative four-form modifications to \eqref{eq:LL4der} consistent with the $\LL$ method, we should not allow such contributions. This in turn implies that the constant $k$ is \emph{exactly} of quartic order in external derivatives.

\subsubsection{Five-derivative terms}
\label{sec:6der-7d}

Furthermore, the reasoning we used in section~\ref{sec:R2+R3} to rule out $\nabla R^2$ terms in the supersymmetry operators in eleven dimensions remains valid in seven. These necessarily add 4-forms to the Lichnerowicz which we have just argued cannot be allowed. However, we are not done at this order. Since we have changed the supersymmetry operators at a lower order, we need to check whether the Lichnerowicz remains consistent at the five-derivative level -- or put in another way, whether the supersymmetry algebra still closes to this order. It does not:
\begin{equation}
\begin{aligned}
\label{eq:6der}
\tilde{D}^mD_m\varepsilon &= -\tfrac14 \mathcal{R}\varepsilon +\tfrac12 k R_{mnpq}R^{mnpq}\varepsilon -\tfrac14 k R_{mnpq}R_{rs}{}^{pq}\gamma^{mnrs} \varepsilon \\&\ph{:=}+  k^2 R_{mnpq}\gamma^{pq} \nabla^n \nabla^r \left(R_{mrst}\gamma^{st} \varepsilon\right).
\end{aligned}
\end{equation}
The last term of the Lichnerowicz does not define a tensor. We must therefore introduce a new correction piece to our operators, which will precisely cancel this last term
\begin{equation}
\begin{aligned}
\Dgen_m\varepsilon &= \LC_m \varepsilon 
     + k \nabla^n \left(R_{mnpq}\gamma^{pq} \varepsilon\right) \\
     &\ph{:= } +4k^2 \tfrac16( \gamma_{mn}-5 g_{mn}) \nabla_p \left( A^{np}{}_{qr} \gamma^{qr}\varepsilon\right)	,\\
\tilde{D}^m \psi_m &= \gamma^{mn}\nabla_m\psi_n 
     + k  R_{mnpq}\gamma^{pq} \nabla^n \psi^m \\
      &\ph{:= } -4k^2 \tilde{A}^{mn}{}_{pq}\gamma^{pq}\nabla_n\psi_m ,
\end{aligned}
\end{equation}
with
\begin{equation}
\begin{aligned}
&A^{mn}{}_{pq}\gamma^{pq}\varepsilon = \nabla^{[m}\nabla_p (R^{n]p}{}_{qr} \gamma^{qr}\varepsilon), \\
&\tilde{A}^{mn}{}_{pq}\gamma^{pq}\varepsilon = R^{p[m}{}_{qr}\nabla_p\nabla^{n]}\gamma^{qr}\varepsilon .
\end{aligned}
\end{equation}
Note that in seven dimensions, $\gamma^{mn}(\gamma_{np} - 5 g_{np})= 6\delta_p^m$. 
These new terms are not of the form $\nabla R^{n}$ which we had considered thus far, they include higher derivatives of the spinor paramenter (also seen in, for example,~\cite{Bergshoeff:1989de}), and clearly do not appear in the eleven-dimensional solution. Their origin is clear, however, as they come multiplied by $k^2$, a factor which is of order $\partial^8$ in external derivatives. In total, we recognise that these new terms are of the same order as $R^7$ corrections in eleven dimensions. If we had simply restricted ourselves to reducing the $R^4$ solution we would never have found these terms, but they are natural, indeed crucial, from the point of view of the internal seven-dimensional supersymmetry.

\subsubsection{Seven-derivative terms}
\label{sec:8der-7d}

Of course, moving to the next order once again breaks the Lichnerowicz, and we have to introduce new corrections, proportional to $k^3$, and which must descend from eleven-dimensional terms of order $R^{10}$. We find
\begin{equation}
\begin{aligned}
\label{eq:Dk-8der}
\Dgen_m\varepsilon &= \LC_m \varepsilon 
     + k \nabla^n \left(R_{mnpq}\gamma^{pq} \varepsilon\right) \\
     &\ph{:= } +4k^2 \tfrac16( \gamma_{mn}-5 g_{mn}) \nabla_p \left( A^{np}{}_{qr} \gamma^{qr}\varepsilon\right)	\\
     &\ph{ := }-32 k^3 \tfrac16 (\gamma_{m}{}^p - 5\delta_m{}^p)\nabla^n\left(B_{npqr}\gamma^{qr}\varepsilon\right),\\
\tilde{D}^m \psi_m &= \gamma^{mn}\nabla_m\psi_n 
     + k  R_{mnpq}\gamma^{pq} \nabla^n \psi^m \\
      &\ph{ := } -4k^2 \tilde{A}^{mn}{}_{pq}\gamma^{pq}\nabla_n\psi_m \\
    &\ph{:= } +32 k^3 \tilde{B}_{mnpq}\gamma^{pq}\nabla^n \psi^m ,
\end{aligned}
\end{equation}
with
\begin{equation}
\begin{aligned}
&B^{mn}{}_{pq}\gamma^{pq}\varepsilon =  \tfrac16 (\gamma^{p[m}+ 5g^{p[m})\nabla^{n]} \nabla^q (A_{pq}{}_{rs}\gamma^{rs}\varepsilon), \\
&\tilde{B}^{mn}{}_{pq}\gamma^{pq}\varepsilon = \tilde{A}^{p[m}{}_{rs} \nabla_p\nabla^{n]}\gamma^{rs}\varepsilon .
\end{aligned}
\end{equation}
This ensures the consistency of the ``$k$ family'' of corrections up to seven internal derivatives. 

At this order we can also return to the $\nabla R^3$ terms and clearly we could effectively just transpose the eleven-dimensional solutions to seven dimensions, performing minimal adjustments, as they only contain zero-forms (and 8-forms, which now vanish identically) by construction. As we saw in section~\ref{sec:11d-r4-min} and appendix~\ref{app:non-min}, there are several such solutions, corresponding to several unfixed coefficients. A certain combination of these seven-dimensional solutions will correspond to taking the eleven-dimensional $R^4$ action as purely internal, so their coefficients will be of zeroth order in external derivatives. Others might correspond to higher-derivative corrections beyond $R^4$ in eleven dimensions which factorise in such a way as to give rise to internal $R^4$ terms.

\subsection{Solutions with $G_2$ holonomy}

We will now make some considerations about the particular case when the seven-dimensional manifold has $G_2$ holonomy.

The assumption of $G_2$ holonomy imposes a large number of simplifications in our formulae. We will not go through the complete solution, but note, for example, that if we take the $G_2$ structure to be defined by a spinor $\varepsilon$, then when evaluating our supersymmetry operators all derivatives of $\varepsilon$ drop out since $\nabla\varepsilon =0$ by $G_2$ holonomy.

Furthermore, we have that $R_{mnpq}\gamma^{pq}\varepsilon = 0$ as well, so the entire  ``$k$ family'' of corrections from the previous subsection vanishes identically. This can be seen either by acting on the supersymmetry variation by another $\nabla$ or introducing $G_2$ invariant 3-form and dual 4-form in terms of the complete basis of seven-dimensional spinors is given by $\{\varepsilon, \gamma^m \varepsilon \}$ \cite{Kaste:2003zd}:
\begin{equation}
\begin{aligned}
\label{eq:gammaphi}
&\gamma_{mn} \varepsilon = \ii \, \phi_{mnp} \gamma^p \varepsilon, \\
&\gamma_{mnp}\varepsilon = \ii \, \phi_{mnp} \, \varepsilon -  *\phi_{mnpq} \gamma^q \varepsilon.
\end{aligned}
\end{equation}
One finds that for $G_2$ holonomy manifolds the Riemann tensor satisfies $R_{mnpq} \phi^{pqr} = 0$, which implies that under the $\bf 28 \rightarrow 21 + 7$ decomposition of a 2-form, the representation $\bf 7$ is missing. This can be written equivalently as $R_{mnpq} (*\phi)^{mn}{}_{rs} = 2 R_{rspq}$.

This brings us to the eight-derivative terms.
In fact, $G_2$ holonomy also implies that the $t_8t_8 R^4$ term vanishes identically~\cite{Gross:1986iv}, so the internal action actually has no corrections at order $R^4$. The equations of motion are still corrected~\cite{Freeman:1986br,Lu:2003ze}, however, and for an external flat space they become simply $\nabla^m \nabla_m Z_{n}{}^n = 0$, where $Z$ is a function of $R^3$ given below. This is then what a $G_2$ Lichnerowicz formula should reproduce.

There exists a well-known correction term to the supersymmetry variation for $G_2$ manifolds. It is usually given in terms of the $G_2$-invariant 3-form $\phi_{mnp}$ as
\begin{equation}
\label{eq:g2-susy}
\delta \psi_m = \nabla_m \varepsilon + \ii \alpha ( \nabla_n Z_{mp})\phi^{npq}\gamma_q  \varepsilon,
\end{equation}
with $\alpha$ some real constant and 
\begin{equation}
\begin{aligned}
\label{eq:Zab}
 Z_{mn} = g\, \epsilon_{m m_1 \dots m_6}\epsilon_{n n_1 \dots n_6}R^{m_1 m_2 n_1 n_2}R^{m_3 m_4 n_3 n_4}R^{m_5 m_6 n_5 n_6}  ,
 \end{aligned}
\end{equation}
is a correction term~\cite{Lu:2003ze} which satisfies $\nabla^m Z_{mn} = 0$  thanks to the Bianchi identity of $R_{mnpq}$. $Z_{mn}$ may be written in terms of our $\SO(7)$ bases as
\begin{equation}
\begin{aligned}
 Z_{mn} = 24 (-\check{W}^1_{mn} + 2 \check{W}^2_{mn} +  \check{W}^3_{mn})  + \tfrac{4}{7} g_{mn}( \check{S}^1 - 2  \check{S}^2 ) .
\end{aligned}
\end{equation}
The precise form of this correction was a crucial part of the analysis of~\cite{Becker:2014rea}, which examined whether the $G_2$ solution remains a valid supersymmetric background to all orders of higher-derivatives corrections.
We will leave for future work a direct comparison of~\eqref{eq:g2-susy} with the reduced $x_5$ solution given in the previous section, but we will remark that just a quick look at the representation theory shows that it is plausible that the two match. First note that under a $G_2$ decomposition, the number of possible terms that can be admitted in $\delta \psi$ is quite small, they are listed in table~\ref{tab:R4-g2}.\footnote{Note that here we do not require that only the antisymmetric $[\nabla,\nabla]$ appears in the $G_2$ Lichnerowicz formula since by using the covariantly constant spinor ($\nabla \varepsilon=0$) we do not need to worry about tensoriality. In fact, as mentioned, the expectation is that the $G_2$  Lichnerowicz  will result in the equation of motion $\nabla^m \nabla_m Z_{n}{}^n = 0$.} 
\begin{table}[h]
\begin{center}
\begin{tabular}{ccc}
\\
Projection of $R^3$& Rep of $G_2$& Multiplicity \\
\hline
\hline
$\dot{I}^i$ & [3,0] & 2 \\
$\dot{K}$ & [1,1] & 1 \\
$\dot{W}^i$ & [2,0] & 3 \\ 
$\dot{S}^i$ & [0,0] & 2 
\end{tabular} 
\caption{Valid embeddings of $\otimes^3 R$ in $\delta\psi$. In each case, the index $i$ runs over the corresponding multiplicity.} 
\label{tab:R4-g2}
\end{center}
\end{table}

We observe that since $[0,2]$ (the representation of the Weyl tensor) is not one of the admissible terms, it is not possible to obtain $R^4$ scalars from Lichnerowicz, which lines up with our expectation that those terms in the action vanish. Now, focusing on the terms that appear in the restricted operator~\eqref{eq:X5reduced}, from the $so(7)$ under $g_2$ branching rules we have the decompositions of the $\check{X}^i$, $[0,2,0] \rightarrow [0,2]+[1,1]+[2,0]$, the $\check{V}^i$, $[1,1,0] \rightarrow [1,1]+[2,0]+[0,1]$ and the $T^i$, $[0,0,2] \rightarrow [2,0]+[1,0]+[0,0]$, though note that the $[1,0]$ and $[0,1]$ components drop out as they do not exist in the tensor product of $R^3$. The $\check{W}^i\in [2,0,0]\rightarrow [2,0]$ do not decompose, nor do the scalars $\check{S}^i$, so in particular $Z_{mn}$ keeps its form. We see that all the terms in the $x_5$ solution mix together in the $[1,1]+[2,0]+[0,0]$ representations, and therefore agreement with~\eqref{eq:g2-susy} will require the $[1,1]$ to vanish and the $[2,0]+[0,0]$ to be precisely $Z_{mn}$. A potential issue is that in eleven dimensions we ignored terms such as the ones of the type $W^i$ and $S^i$ since they did not contribute to Lichnerowicz (they do not affect the bosonic action), yet they might be required here in order to obtain a precise match at the level of the operators.

Finally, we remark that the calculation of~\cite{Lu:2003ze} which first derived~\eqref{eq:g2-susy} relies on this operator satisfying an integrability condition precisely of the type that first led us to consider more general Lichnerowicz formulae, so in that sense the problem has already been solved.

\subsection{Comments on compactifications}
\label{ssec:dis}

We shall now comment on the relation between the results of subsection \ref{ssec:ML} and the general results of section \ref{sec:high}. The seven-derivative contributions to the covariant derivative $D_a$ are clearly related, but while for the eleven-dimensional operator these are the first higher-derivative terms, its seven-dimensional counterpart has also three- and five-derivative terms. Where do these come from?

Let us start with lowest order, i.e. the three-derivative terms in \eqref{eq:3der}. If one thinks of $k$ not as a numerical coefficient, but a combination quadratic in external (four-dimensional) Riemann tensors, then one can find factorised terms in \eqref{eq:Dmin} and \eqref{eq:tDmin} that can yield \eqref{eq:3der} upon breaking eleven-dimensional Lorentz invariance.

The $y_2$ and $x_5$ families behave rather differently. $\tfrac12 x_5(  R_{46} -4   A_7)$ appearing in \eqref{eq:LL11D} does not have any factorised terms and hence its reduction would lead directly to $k=0$.  The reducible part of the $y_2$ family that yields a non-trivial $k$ contribution is $- \tfrac12 y_2 R_{44} $.

Let us also observe that if one precipitates and uses \eqref{eq:gammaphi} together with $G_2$ self-duality relations for the Riemann curvature, one find that the two terms proportional to $k$ cancel out in \eqref{eq:LL4der}. In our approach, we keep the zero-form and the four-form separately and add the latter to the action after completing it to a top-form by wedging with $C$.

In the reduction on a $G_2$ holonomy manifold $X$, the relevant part of the eleven-dimensional action to be integrated over $X$ is given by
\begin{equation}
\begin{aligned}
\label{eq:red4der}
\tfrac12  (R_{abcd}R^{abcd})^2 -\tfrac14 C \wedge \tr R^2 \wedge \tr R^2 \mapsto +\tfrac12 E_4\int_X \phi \wedge \tr R^2  -\tfrac14 \tr R^2 u_i \int_X \omega_3^i \tr R^2  .
\end{aligned}
\end{equation}
Here $E_4$ is the four-dimensional Euler density which (up to Ricci terms) is the same as the Riemann tensor squared. In the first term on the right-hand side we have used that $R_{mnpq}R^{mnpq} = \tfrac12 R_{mnpq}R_{rs}{}^{pq} (*\phi)^{mnrs} =  *(\phi \wedge \tr R^2)$. Finally $\omega^i \in H^3(X)$ and $C_3 = u_i \, \omega_3^i$, with $i=1, \cdots, b_3(X)$. We may decompose similarly $\phi = t_i \, \omega_3^i$. In fact $u_i$ and $t_i$ form the scalar sector of $b_3$ chiral superfields of the $N=1$ theory in four dimensions. We end up with a four-derivative contribution to the $N=1$ effective theory:
\begin{equation}
\begin{aligned}
\label{eq:4D}
\mathcal{L}_{N=1} \sim \alpha_i \Big(  \tfrac12 u_i  E_4 + t_i \tr R^2 \Big),
\end{aligned}
\end{equation}
with 
\begin{equation}
\begin{aligned}
\alpha^i = \int_X \omega_3^i \wedge \tr R^2 .
\end{aligned}
\end{equation}

For $X = {\tilde X} \times S^1$, where $\tilde X$ is a Calabi-Yau threefold, one can recognise the familiar one-loop $R^2$ couplings in $N=2$ theories, where now in the internal six-dimensional integral the $\omega^i$ are replaced by the forms in the $H^{(1,1)}({\tilde X})$, and $u_i + \ii \, t_i$ is the complex scalar in the $N=2$ vector multiplets \cite{Antoniadis:1997eg}.

As mentioned earlier and further elaborated below, the  terms in action with $k$ factors should not receive corrections from higher-derivative (higher string loop terms) and hence we do not expect the coupling \eqref{eq:4D} to receive further perturbative corrections.
We may remark that the moduli spaces of $G_2$ compactifications are not factorised and have (complex) dimension $b_3(X) + b_2(X)$.\footnote{We discussed here the $b_3(X)$ chiral multiplets made of deformations of the metric and the scalar modes coming from $C_3$. The latter also yields $b_2(X)$ vector fields, which make the bosonic part of vector multiplets.} It is curious that the higher-derivative couplings make use of $b_3(X)$ topological numbers $\alpha^i$ and distinguish between the two sectors.

We can turn to the next order -- five derivative terms in the operators. Note that these are designed to cancel non-tensional terms in \eqref{eq:6der}. However, they contain a factor of $k^2$ on top of two Riemann tensors and two $\nabla$, and so from the eleven-dimensional point of view are order $R^7$. Hence the additional contributions to the supersymmetry operators are reductions of the next order  terms in eleven dimensions, i.e. 13 derivative terms. Similarly, at the next order of internal derivatives we have terms $\sim k^3$, and these come from the reduction of 19 derivative corrections to eleven-dimensional supersymmetry. Note that we are only probing the fraction of the higher (than seven) derivative terms in supersymmetry whose purpose is to cancel unwanted non-tensorial contributions from cross-terms of lower-derivative contributions. By design the action itself stays order $\sim R^4$. Notice, however, that the seven-dimensional $\LL$ method seems to be giving information about terms that are higher order in the eleven-dimensional sense.

We may finally comment on factorisability properties of higher derivative terms. In \cite{Antoniadis:2003sw} the terms that allowed to factor out a single Ricci scalar were discussed. Here we completely ignore the Ricci terms, so we cannot further comment on corrections of that type. However, our seven-dimensional calculation would appear to rule out certain other types of terms in eleven dimensions. Consider, for example, an order $R^7$ correction to the action that factorises as $R^3 \cdot R_{44}$, where $R^3$ is some linear combination of the $S^i$ given in appendix~\ref{app:projR4} and $R_{44}$ was given in~\eqref{eq:R4-scalars}. Upon decomposing to seven dimensions, this would lead to a term $\tilde{k}(R^{\text{7d}})^3$ in the reduced action, with $\tilde{k}= R^{\text{4d}}_{44}= (R_{\mu\nu\lambda\rho}R^{\mu\nu\lambda\rho})^2$. This would be in contradiction with our Lichnerowicz calculation in section~\ref{sec:8der-7d} which disallowed $R^3$ terms in seven dimensions. Thus, we are lead to conclude that corrections of the type $R^3\cdot [\text{anything nonzero in 4d}]$ appear to be ruled out in eleven dimensions. On the other hand, a pure Riemann term like $R^3 \cdot (\epsilon_8\epsilon_8R^4)$ is possible. Similar considerations also imply that (beyond order $R^4$) terms that factorise as $R^2\cdot [\text{anything nonzero in 4d}]$ are likewise ruled out.

\section{Future directions}
\label{sec:future}

The effects induced by higher-derivative corrections play an important role in many lower dimensional theories arising from compactifications. It may be useful, even if more laborious, to work out a full reduction of the eleven-dimensional $\LL$ formula, rather than study a lower dimensional ``descendant", as we did here. An intriguing aspect of studying the formula on a product of internal and external spaces is that different derivative orders get mixed on a lower dimensional component, allowing for glimpses into further corrections to supersymmetry operators. There are still open problems concerning the eleven-derivative covariant couplings, and we conclude by mentioning two venues of possible progress.

\subsection{Adding fluxes}

We began by showing that the classical supersymmetry operators with flux obey a Lichnerowicz relation that reproduces the classical action. However, when we moved to the higher-order corrections we set $G=0$ in order to simplify the computation. A clear next step is to restore those terms and obtain the flux completion of $R^4$. 

The issue of computing  the full set of $G$ flux contributions to the eight derivative corrections is a long standing one.  Progress has been made in string theory, and it can be shown that at one loop most of the NS sector contributions are captured by introducing a connection with torsion $\omega^{\mbox{\tiny LC}} \, \rightarrow \omega^{\mbox{\tiny LC}} + H$. There are, however, additional ambiguities associated with lifting to eleven-dimensions, and replacing $H$ by $G$ \cite{Liu:2013dna}.

In the context of the M-theoretic Lichnerowicz method, the computation of the flux terms should proceed in a straightforward iterative manner. Firstly, promoting the leading term $\nabla_a \rightarrow \nabla_a^G$ in the supersymmetry operator $D_a$ will immediately break the tensoriality of $\tilde{D}^aD_a\varepsilon$ and will require adjustments to the existing higher-derivative terms -- for example, all the $\nabla_a$ in the correction terms of $\tilde{D}_a$ will likewise have to be replaced by $\nabla_a^G$. Additionally, we neglected Ricci terms but now they become proportional to $G^2$ by the  equations of motion so they cannot be ignored. Instead, we have that factors of $\gamma^a[\nabla_a^G,\nabla_b^G]$ and $(\tilde{\nabla}^G)^a\nabla^G_a$ give the combinations of covariant derivatives that vanish on-shell. Finally, it will be necessary to impose the constraint that the M-theory Lichnerowicz define just a scalar and an 8-form. This will require the addition of new terms to the corrected operators to cancel the other $p$-forms, which should be easier to do if one proceeds order by order in powers of the $G$ flux. 

Note that at four-derivative order, the couplings in the seven-dimensional theory with 16 supercharges including four-form flux are known without ambiguities \cite{Liu:2013dna}.  This case should provide a good test for completing  the seven-dimensional $\LL$ formula  with $G$.

\subsection{Towards $R^7$}

The M-theoretic Lichnerowicz computed from the operators we defined in section~\ref{sec:11d-r4-min} $\tilde{D}^a D_a \varepsilon$ only results in a scalar and a 3-form up to order $R^4$. If we include the higher order terms in the computation this will fail, and the form of this failure is clear -- if we schematically write the corrections $D = \nabla + \nabla R^3$ and $\tilde{D}= \nabla + R^3 \nabla$, then the full $\tilde{D}D$ will include $R^3 \nabla \nabla R^3 $, which will even involve non-tensor terms. These terms are of the same order as $R^7$, so we are led to conclude that no new corrections will be needed at order $R^5$ or $R^6$, but are \emph{necessary} at $R^7$. This is consistent with the fact that in the strong coupling eleven-dimensional limit \cite{Witten:1995ex}, the only surviving terms  will be of order $R^{3l+1}$ for loop order $l$. 

In principle, the Lichnerowicz procedure should allow us to deduce what is the form of these corrections. However, computationally it is significantly more difficult to generate the necessary tensor products of seven Riemann tensors, and, in addition, it is expected that at this order there might be terms involving explicit connections $\nabla$ in the action, further complicating matters. Nevertheless, since we can easily  compute  the `problem' terms explicitly in the Lichnerowicz equation, it seems hopeful that one can cancel them step-by-step by suitable modifications of the $D$ and $\tilde{D}$ operators, just like we did in the simpler seven-dimensional case in section~\ref{sec:6der-7d}. As an example, consider just the $X^1$ part of the $y_2$ solution for the operators~\eqref{eq:Dmin}, i.e.
\begin{equation}
\begin{aligned}
D_a \varepsilon &= \nabla_a \varepsilon - y_2 \nabla^b ( X^1_{abcd}\gamma^{cd}\varepsilon ) +\dots \\
&= \nabla_a \varepsilon - y_2 \nabla^b (R^2 R_{abcd}\gamma^{cd}\varepsilon ) +\dots \\
\tilde{D}^a \psi_a &= \gamma^{ab}\nabla_a \psi_b - y_2  X^1_{abcd}\gamma^{cd}\nabla^b \psi^a +\dots \\
&= \gamma^{ab}\nabla_a \psi_b - y_2 R^2 R_{abcd}\gamma^{cd} \nabla^b \psi^a+\dots \\
\end{aligned}
\end{equation}
which gives in the M-theory Lichnerowicz
\begin{equation}
\begin{aligned}
\label{eq:higher-order11L}
\tilde{D}^a D_a \varepsilon = \dots + (y_2)^2   R^2 R^a{}_{bcd}\gamma^{cd} \nabla^b \nabla^e (R^2 R_{aefg}\gamma^{fg}\varepsilon ) +\dots 
\end{aligned}
\end{equation}
This term is analogous to the one we found in seven dimensions, and so can be cancelled in a similar manner. We therefore expect that there will be a correction to the supersymmetry operators at 13 derivatives given by
\begin{equation}
\begin{aligned}
D_a \varepsilon &= \nabla_a \varepsilon  - y_2 \nabla^b (R^2 R_{abcd}\gamma^{cd}\varepsilon ) \\ &\ph{:=}+4 (y_2)^2 \tfrac{1}{10} (\gamma_{ab}- 9g_{ab}) \nabla_c \left( R^2\nabla^{[b}\nabla_d ( R^{c]d}{}_{ef} R^2\gamma^{ef}\varepsilon )\right) +\dots \\
\tilde{D}^a \psi_a &= \gamma^{ab}\nabla_a \psi_b- y_2 R^2 R_{abcd}\gamma^{cd} \nabla^b \psi^a \\
&\ph{:=}-4 (y_2)^2   R^{c [a}{}_{ef} R^2 \nabla_c \nabla^{b]}  \gamma^{ef}(R^2 \nabla_b \psi_a ) +\dots \\
\end{aligned}
\end{equation}
Since the effect of this particular correction is to precisely cancel the non-tensorial piece in~\eqref{eq:higher-order11L}, it does not generate new contributions to the action of order $R^7$.

\begin{acknowledgments}

We would like to thank K. Becker,  D. Robbins, A. Royston, P. Vanhove for useful discussions.
A.~C.~has been supported by the Laboratoire d'Excellence CARMIN. 
A.~C.~also thanks the Program on the Mathematics of String Theory at Institut Henri Poincar\'e, Paris for hospitality during the completion of this work. The work of R.~M.   is supported in part by the Agence Nationale de la Recherche under the grant 12-BS05-003-01.
\end{acknowledgments}

\appendix

\section{Projectors for $R^3$} 

\subsection{In eleven dimensions}
\label{app:projR4}

Here we present bases for projections of tensor products of Weyl tensors into irreducible representations of $SO(10,1)$. Throughout, same letter indices which are free are assumed to be antisymmetrised with unit weight (contracted indices have no such assumption). We will be considering the irreducible representations listed in table~\ref{tab:R4a} which are relevant for the Lichnerowicz calculation. For convenience, these are reproduced in table~\ref{tab:appR4a}.

\begin{table}[h]
\begin{center}
\begin{tabular}{ccc}
\\
Projection of $R^3$ & Rep  of $SO(10,1)$ & Multiplicity \\
\hline
\hline
$X^i$ & [0,2,0,0,0] & 8 \\
$W^i$ & [2,0,0,0,0] & 3 \\
$S^i$ & [0,0,0,0,0] & 2 \\ \hline
$Y^i$ & [0,1,0,0,2] & 2 \\
$V^i$ & [1,0,0,0,2] & 2 \\ 
$T^i$ & [0,0,0,1,0] & 3 \\ \hline
$Z^i$ & [0,1,0,1,0] & 3 \\
$U^i$ & [1,0,1,0,0] & 3 \\ 
\end{tabular} 
\caption{Relevant projections of $\otimes^3 R$ in eleven dimensions. In each case, the index $i$ runs over the corresponding multiplicity.} 
\label{tab:appR4a}
\end{center}
\end{table}

The four-index $X^i$ terms contain two pairs of antisymmetric indices which are symmetric under exchange, and are fully traceless:
\begin{equation*}
\begin{aligned}
X^1_{a_1 a_2 b_1 b_2} &= R_{a_1 a_2 b_1 b_2} R_{c_1 c_2 d_1 d_2} R_{c_1 c_2 d_1 d_2},\\
X^2_{a_1 a_2 b_1 b_2} &= \tfrac12 R_{a_1 a_2 b_1  c_1} R_{b_2 c_2 d_1 d_2} R_{c_1 c_2 d_1 d_2}
+  \tfrac12 R_{b_1 b_2 a_1  c_1} R_{a_2 c_2 d_1 d_2} R_{c_1 c_2 d_1 d_2}\\&\ph{:=}
+ \tfrac29 g_{a_2 b_1} R_{b_2 d_1 a_1 d_2} R_{d_1 c_1 c_2 e_1} R_{d_2 c_1 c_2 e_1},\\
X^3_{a_1 a_2 b_1 b_2} &= \tfrac23 R_{a_1 a_2 c_1 c_2} R_{b_1 b_2 d_1 d_2} R_{c_1 c_2 d_1 d_2}
- \tfrac23  R_{a_1 b_1 c_1 c_2} R_{b_2 a_2  d_1 d_2} R_{c_1 c_2 d_1 d_2}\\&\ph{:=}
+ \tfrac49 g_{a_2 b_1} R_{b_2 e_1 d_1 d_2} R_{d_1 d_2 c_1 c_2} R_{c_1 c_2 a_1 e_1}\\&\ph{:=}
 -\tfrac{1}{45} g_{a_2 b_1} g_{a_1 b_2} R_{d_1 d_2 e_1 e_2} R_{e_1 e_2 c_1 c_2} R_{c_1 c_2 d_1 d_2} ,\\
X^4_{a_1 a_2 b_1 b_2} &= \tfrac23 R_{a_1 c_1 b_1 c_2} R_{a_2 d_1 b_2 d_2} R_{c_1 d_1 c_2 d_2}
+ \tfrac13  R_{a_1 c_1 b_1 c_2} R_{a_2 d_1 b_2 d_2 } R_{c_1 d_2 c_2 d_1}\\&\ph{:=}
+ \tfrac{1}{24} R_{a_1 a_2 c_1 c_2} R_{b_1 b_2 d_1 d_2} R_{c_1 c_2 d_1 d_2}\\&\ph{:=}
+ \tfrac{1}{18} g_{a_2 b_1} R_{b_2 e_1 d_1 d_2} R_{d_1 d_2 c_1 c_2} R_{c_1 c_2 a_1 e_1}  -\tfrac29 g_{a_2 b_1} R_{b_2 d_1 e_1 d_2} R_{d_1 c_1 d_2 c_2} R_{c_1 a_1 c_2 e_1}\\&\ph{:=}
 -\tfrac{1}{360} g_{a_2 b_1} g_{a_1 b_2} R_{d_1 d_2 b3 b4} R_{b3 b4 c_1 c_2} R_{c_1 c_2 d_1 d_2} + \tfrac{1}{90} g_{a_2 b_1} g_{a_1 b_2} R_{d_1  e_1 d_2 e_2} R_{e_1 c_1 e_2 c_2} R_{c_1 d_1 c_2 d_2},\\
X^5_{a_1 a_2 b_1 b_2} &= \tfrac13 R_{a_1 a_2 c_1 c_2} R_{b_1 c_1 d_1 d_2} R_{b_2 c_2 d_1 d_2} 
+ \tfrac13  R_{b_1 b_2 c_1 c_2} R_{a_1 c_1 d_1 d_2} R_{a_2 c_2 d_1 d_2}\\&\ph{:=}
- \tfrac23   R_{a_1 b_1 c_1 c_2} R_{b_2 c_1 d_1 d_2} R_{a_2 c_2 d_1 d_2}\\&\ph{:=}
 + \tfrac29  g_{a_2 b_1} R_{b_2 e_1 d_1 d_2} R_{d_1 d_2 c_1 c_2} R_{c_1 c_2 a_1 e_1}\\ &\ph{:=}
 -\tfrac{1}{90} g_{a_2 b_1} g_{a_1 b_2} R_{d_1 d_2 e_1 e_2} R_{e_1 e_2 c_1 c_2} R_{c_1 c_2 d_1 d_2} ,\\
\end{aligned}
\end{equation*}

\begin{equation*}
\begin{aligned}
X^6_{a_1 a_2 b_1 b_2} &= \tfrac23 R_{a_1 c_1 b_1 c_2} R_{a_2 d_1 c_1 d_2} R_{b_2 d_1 c_2 d_2}
+ \tfrac13  R_{a_1 c_2 b_1 c_1} R_{a_2 d_1 c_1 d_2} R_{b_2 d_1 c_2 d_2} \\&	\ph{:=}	
+ \tfrac{1}{12} R_{a_1 a_2 c_1 c_2} R_{b_1 d_1 c_1 d_2} R_{b_2 d_1 c_2 d_2} 
+ \tfrac{1}{12} R_{b_1 b_2 c_1 c_2} R_{a_1 d_1 c_1 d_2} R_{a_2 d_1 c_2 d_2}\\&\ph{:=}
 -\tfrac{1}{18}  g_{a_2 b_1}  R_{b_2 e_1 d_1 d_2} R_{d_1 d_2 c_1 c_2} R_{c_1 c_2 a_1 e_1} +\tfrac29 g_{a_2 b_1}  R_{b_2 d_1 e_1 d_2} R_{d_1 c_1 d_2 c_2} R_{c_1 a_1 c_2 e_1}  \\&\ph{:=}+\tfrac19 g_{a_2 b_1}  R_{b_2 d_1 a_1 d_2} R_{d_1 c_1 c_2 e_1} R_{d_2 c_1 c_2 e_1}\\&\ph{:=}
  +\tfrac{1}{360} g_{a_2 b_1} g_{a_1 b_2} R_{d_1 d_2 e_1 e_2} R_{e_1 e_2 c_1 c_2} R_{c_1 c_2 d_1 d_2} -\tfrac{1}{90} g_{a_2 b_1} g_{a_1 b_2}R_{d_1  e_1 d_2 e_2} R_{e_1 c_1 e_2 c_2} R_{c_1 d_1 c_2 d_2},\\
X^7_{a_1 a_2 b_1 b_2} &= \tfrac23 R_{a_1 c_1 b_1 c_2} R_{a_2 c_2 d_1 d_2} R_{b_2 c_1 d_1 d_2} 
+ \tfrac13 R_{a_1 c_2 b_1 c_1}  R_{a_2 c_2 d_1 d_2} R_{b_2 c_1 d_1 d_2}\\&\ph{:=}
- \tfrac{1}{12} R_{a_1 a_2 c_1 c_2} R_{b_1 c_1 d_1 d_2} R_{b_2 c_2 d_1 d_2}
- \tfrac{1}{12} R_{b_1 b_2 c_1 c_2} R_{a_1 c_1 d_1 d_2} R_{a_2 c_2 d_1 d_2}\\&\ph{:=}
-\tfrac29  g_{a_2 b_1} R_{b_2 e_1 d_1 d_2} R_{d_1 d_2 c_1 c_2} R_{c_1 c_2 a_1 e_1}  + \tfrac19 g_{a_2 b_1} R_{b_2 d_1 a_1 d_2} R_{d_1 c_1 c_2 e_1} R_{d_2 c_1 c_2 e_1} \\&\ph{:=}
+ \tfrac{1}{90} g_{a_2 b_1} g_{a_1 b_2} R_{d_1 d_2 e_1 e_2} R_{e_1 e_24 c_1 c_2} R_{c_1 c_2 d_1 d_2} ,\\
X^8_{a_1 a_2 b_1 b_2} &= \tfrac13 R_{a_1 a_2 c_1 c_2} R_{b_1 d_1 c_1 d_2} R_{b_2 d_1 c_2 d_2} 
+ \tfrac13 R_{b_1 b_2 c_1 c_2} R_{a_1 d_1 c_1 d_2} R_{a_2 d_1 c_2 d_2}\\&\ph{:=}
- \tfrac23  R_{a_1 b_1 c_1 c_2} R_{b_2 d_1 c_1 d_2} R_{a_2 d_1 c_2 d_2}\\&\ph{:=}
+ \tfrac49 g_{a_2 b_1} R_{b_2 d_1 e_1 d_2} R_{d_1 c_1 d_2 c_2} R_{c_1 a_1 c_2 e_1} \\&\ph{:=}
-\tfrac{1}{45} g_{a_2 b_1} g_{a_1 b_2} R_{d_1  e_1 d_2 e_2} R_{e_1 c_1 e_2 c_2} R_{c_1 d_1 c_2 d_2} .
\end{aligned}
\end{equation*}

The two-index $W^i$ are symmetric traceless:
\begin{equation*}
\begin{aligned}
W^1_{a_1 b_1} &= R_{a_1 d_1 e_1 e_2} R_{e_1 e_2 c_1 c_2}R_{ c_1 c_2 b_1 d_1} -\tfrac{1}{11} g_{a_1 b_1} R_{e_1 e_2 d_2 b_2} R_{d_2 b_2 c_1 c_2}R_{ c_1 c_2 e_1 e_2},\\
W^2_{a_1 b_1} &= R_{a_1 e_1 d_1 e_2} R_{e_1 c_1 e_2 c_2}R_{ c_1 b_1 c_2 d_1} -\tfrac{1}{11} g_{a_1 b_1} R_{e_1 a_2 e_2 b_2} R_{a_2 c_1 b_2 c_2}R_{ c_1 e_1 c_2 e_2},\\
W^3_{a_1 b_1} &=R_{a_1 e_1 b_1 c_1} R_{e_1 e_2 d_1 d_2}R_{ c_1 e_2 d_1 d_2}.
\end{aligned}
\end{equation*}

The $S^i$ are scalars:
\begin{equation*}
\begin{aligned}
S^1&= R_{a_1 a_2 b_1 b_2} R_{b_1 b_2 c_1 c_2}R_{ c_1 c_2 a_1 a_2},\\
S^2&= R_{a_1 b_1 a_2 b_2} R_{b_1 c_1 b_2 c_2}R_{ c_1 a_1 c_2 a_2}.
\end{aligned}
\end{equation*}

The eight-index $Y^i$ have a pair of antisymmetric indices, another set of six antisymmetric indices, vanish if any seven indices are antisymmetrised, and are traceless:
\begin{equation*}
\begin{aligned}
Y^1_{a_1 a_2 b_1 \dots b_6} &= R_{a_1 a_2 b_1 b_2} R_{b_3 b_4 d_1 d_2} R_{b_5 b_6 d_1 d_2}\\
&\ph{:=}-  \tfrac85 g_{a_1 b_1} R_{c_1 a_2 b_2 b_3} R_{c_1 b_4 d_1 d_2} R_{b_5 b_6 d_1 d_2}\\
&\ph{:=}+   \tfrac{2}{15} g_{a_1 b_1} g_{a_2 b_2} R_{c_1 c_2 b_3 b_4} R_{d_1 d_2 b_5 b_6} R_{c_1 c_2 d_1 d_2}
-   \tfrac{4}{15} g_{a_1 b_1} g_{a_2 b_2} R_{c_1 c_2 b_3 b_4} R_{c_1 b_5 d_1 d_2} R_{c_2 b_6 d_1 d_2},\\
Y^2_{a_1 a_2 b_1 \dots b_6}  &= R_{a_1 c_1 b_1 b_2} R_{a_2 c_2 b_3 b_4} R_{b_5 b_6 c_1 c_2}\\
&\ph{:=}-  \tfrac25 g_{a_1 b_1}  R_{c_1 a_2 b_2 b_3} R_{c_1 b_4 d_1 d_2} R_{b_5 b_6 d_1 d_2}
-  \tfrac45 g_{a_1 b_1} R_{c_1 d_1 b_3 b_4} R_{a_2 d_2 c_1 b_2} R_{d_1 d_2 b_5 b_6}\\
&\ph{:=}+  \tfrac{1}{30} g_{a_1 b_1} g_{a_2 b_2} R_{c_1 c_2 b_3 b_4} R_{d_1 d_2 b_5 b_6} R_{c_1 c_2 d_1 d_2}
-   \tfrac{2}{15}  g_{a_1 b_1} g_{a_2 b_2} R_{c_1 c_2 b_3 b_4} R_{c_1 b_5 d_1 d_2} R_{c_2 b_6 d_1 d_2}\\
&\ph{:=}+   \tfrac{2}{15}  g_{a_1 b_1} g_{a_2 b_2} R_{d_1 d_2 b_3 b_4} R_{c_1 d_1 c_2 b_5} R_{c_2 d_2 c_1 b_6}.
\end{aligned}
\end{equation*}

The six-index $V^i$ have a set of five antisymmetric indices, vanish if the six indices are antisymmetrised, and are traceless:
\begin{equation*}
\begin{aligned}
V^1_{a_2 b_1 \dots b_5}  &= R_{c_1 a_2 b_2 b_3} R_{c_1 b_4 d_1 d_2} R_{b_5 b_1 d_1 d_2}\\
&\ph{:=}- \tfrac17  g_{a_2 b_2} R_{c_1 c_2 b_5 b_1} R_{d_1 d_2 b_3 b_4} R_{c_1 c_2 d_1 d_2}
+ \tfrac27   g_{a_2 b_2} R_{c_1 c_2 b_5 b_1} R_{d_1 d_2 c_1 b_3} R_{d_1 d_2 c_2 b_4},\\
V^2_{a_2 b_1 \dots b_5} &=  R_{c_1 d_1 b_3 b_4} R_{a_2 d_2 c_1 b_2} R_{d_1 d_2 b_5 b_1}\\
 &\ph{:=}- \tfrac17 g_{a_2 b_2} R_{c_1 c_2 b_5 b_1} R_{d_1 d_2 c_1 b_3} R_{d_1 d_2 c_2 b_4}
 - \tfrac27   g_{a_2 b_2} R_{d_1 d_2 b_5 b_1} R_{c_1 d_1 c_2 b_3} R_{c_2 d_2 c_1 b_4}.
\end{aligned}
\end{equation*}

The $T^i$ are 4-forms:
\begin{equation*}
\begin{aligned}
T^1_{b_1 \dots b_4} &= R_{c_1 c_2 b_1 b_2} R_{d_1 d_2 b_3 b_4} R_{c_1 c_2 d_1 d_2},\\
T^2_{b_1 \dots b_4} &= R_{c_1 c_2 b_1 b_2} R_{d_1 d_2 c_1 b_3} R_{d_1 d_2 c_2 b_4},\\
T^3_{b_1 \dots b_4} &= R_{d_1 d_2 b_1 b_2} R_{c_1 d_1 c_2 b_3} R_{c_2 d_2 c_1 b_4}.\\
\end{aligned}
\end{equation*}

The six-index $Z^i$ have a pair of antisymmetric indices, another set of four antisymmetric indices, vanish if any five indices are antisymmetrised, and are traceless:
\begin{equation*}
\begin{aligned}
Z^1_{a_1 a_2 b_1 \dots b_4} &= \tfrac12 R_{a_1 a_2 d_1 b_1} R_{c_1 c_2 d_1 b_2} R_{c_1 c_2 b_3 b_4}
+ \tfrac12 R_{b_2 b_1 d_1 a_2} R_{c_1 c_2 d_1 a_1} R_{c_1 c_2 b_3 b_4}\\
&\ph{:=}+  \tfrac17 g_{a_2 b_4} R_{a_1 c_1 b_1 b_2} R_{d_1 d_2 c_2 c_1} R_{d_1 d_2 c_2 b_3}
+\tfrac17 g_{a_2 b_4} R_{c_1 c_2 b_1 b_2} R_{c_1 b_3 d_1 d_2}R_{d_1 d_2 c_2 a_1}\\
&\ph{:=}+\tfrac17 g_{a_2 b_4} R_{c_1 c_2 a_1 b_2} R_{c_1 b_3 d_1 d_2}R_{d_1 d_2 c_2 b_1},\\
Z^2_{a_1 a_2 b_1 \dots b_4} &= \tfrac12 R_{a_1 c_1 d_1 b_1} R_{c_1 c_2 d_1 b_2} R_{a_2 c_2 b_3 b_4}
+  \tfrac12 R_{a_1 d_1 c_1 b_1} R_{c_1 c_2 d_1 b_2} R_{a_2 c_2 b_3 b_4}\\
&\ph{:=}-\tfrac{3}{28} g_{a_2 b_4} R_{a_1 c_1 b_1 b_2} R_{d_1 d_2 c_2 c_1} R_{d_1 d_2 c_2 b_3}
-\tfrac{1}{28} g_{a_2 b_4} R_{c_1 c_2 b_1 b_2} R_{c_1 b_3 d_1 d_2}R_{d_1 d_2 c_2 a_1}\\
&\ph{:=}-\tfrac{1}{28} g_{a_2 b_4} R_{c_1 c_2 a_1 b_2} R_{c_1 b_3 d_1 d_2}R_{d_1 d_2 c_2 b_1}
+\tfrac{1}{7} g_{a_2 b_4}  R_{c_1 c_2 b_1 b_2} R_{c_1 d_1 b_3 d_2}R_{d_1 c_2 d_2 a_1}\\
&\ph{:=}+\tfrac{1}{7} g_{a_2 b_4}  R_{c_1 c_2 a_1 b_2} R_{c_1 d_1 b_3 d_2}R_{d_1 c_2 d_2 b_1},\\
Z^3_{a_1 a_2 b_1 \dots b_4} &= \tfrac12 R_{a_1 d_1 c_1 c_2} R_{c_1 c_2 b_1 b_2} R_{a_2 d_1 b_3 b_4}
+  \tfrac12 R_{b_1 d_1 c_1 c_2} R_{c_1 c_2 a_1 b_2} R_{a_2 d_1 b_3 b_4}\\
&\ph{:=}-\tfrac{3}{14} g_{a_2 b_4} R_{a_1 c_1 b_1 b_2} R_{d_1 d_2 c_2 c_1} R_{d_1 d_2 c_2 b_3}
-\tfrac{1}{14} g_{a_2 b_4} R_{c_1 c_2 b_1 b_2} R_{c_1 b_3 d_1 d_2}R_{d_1 d_2 c_2 a_1}\\
&\ph{:=}-\tfrac{1}{14} g_{a_2 b_4} R_{c_1 c_2 a_1 b_2} R_{c_1 b_3 d_1 d_2}R_{d_1 d_2 c_2 b_1}.
\end{aligned}
\end{equation*}

The four-index $U^i$ have a set of three antisymmetric indices, vanish if the four indices are antisymmetrised, and are traceless:
\begin{equation*}
\begin{aligned}
U^1_{a_1 b_1 b_2 b_3} &= R_{a_1 c_1 b_1 b_2} R_{d_1 d_2 c_2 c_1} R_{d_1 d_2 c_2 b_3},\\
U^2_{a_1 b_1 b_2 b_3}  &= \tfrac12 R_{c_1 c_2 b_1 b_2} R_{c_1 d_1 b_3 d_2}R_{d_1 c_2 d_2 a_1}
+\tfrac12 R_{c_1 c_2 a_1 b_2} R_{c_1 d_1 b_3 d_2}R_{d_1 c_2 d_2 b_1},\\
U^3_{a_1 b_1 b_2 b_3} &= \tfrac12 R_{c_1 c_2 b_1 b_2} R_{c_1 b_3 d_1 d_2}R_{d_1 d_2 c_2 a_1}
+\tfrac12 R_{c_1 c_2 a_1 b_2} R_{c_1 b_3 d_1 d_2}R_{d_1 d_2 c_2 b_1}.
\end{aligned}
\end{equation*}

\subsection{In seven dimensions}
\label{app:projR4-7}

In principal, the seven-dimensional supersymmetry operator can contain any of the $SO(7)$ representations listed in table~\ref{tab:R4-7}. However, we write explicitly only the projections that will be relevant for a reduction of the $x_5$ solution in eleven-dimensions from section~\ref{sec:11d-r4-min}.

\begin{table}[htb]
\begin{center}
\begin{tabular}{ccc}
\\
Projection of $R^3$& Rep of $SO(7)$& Multiplicity \\
\hline
\hline
$\check{X}^i$ & [0,2,0] & 7 \\
$\check{W}^i$ & [2,0,0] & 3 \\
$\check{S}^i$ & [0,0,0] & 2 \\ \hline
$\check{V}^i$ & [1,1,0] & 2 \\ 
$\check{T}^i$ & [0,0,2] & 3 \\ \hline
$\check{Z}^i$ & [0,1,2] & 3 \\
$\check{U}^i$ & [1,0,2] & 3 \\
\hline
\hline 
$\check{L}^i$ & [2,1,0] & 3 \\
$\check{M}^i$  & [2,0,2] & 6 
\end{tabular} \\
\caption{Valid embeddings of $\otimes^3 R$ in $\delta\psi$. In each case, the index $i$ runs over the corresponding multiplicity.} 
\label{tab:R4-7}
\end{center}
\end{table}
Indices $m,n\dots$ run from 1 to 7, and, as before, any free indices with the same letter are assumed to be antisymmetrised. The metric $g$ is now that of the internal seven-dimensional space.

We will consider two $\check{X}^i$ terms. They contain two pairs of antisymmetric indices which are symmetric under exchange, and are fully traceless:
\begin{equation*}
\begin{aligned}
\check{X}^5_{m_1 m_2 n_1 n_2} &= \tfrac13 R_{m_1 m_2 p_1 p_2} R_{n_1 p_1 q_1 q_2} R_{n_2 p_2 q_1 q_2} 
+ \tfrac13  R_{n_1 n_2 p_1 p_2} R_{m_1 p_1 q_1 q_2} R_{m_2 p_2 q_1 q_2} \\&\ph{:=}
- \tfrac23   R_{m_1 n_1 p_1 p_2} R_{n_2 p_1 q_1 q_2} R_{m_2 p_2 q_1 q_2}\\&\ph{:=}
 + \tfrac25  g_{m_2 n_1} R_{n_2 r_1 q_1 q_2} R_{q_1 q_2 p_1 p_2} R_{p_1 p_2 m_1 r_1}\\ &\ph{:=}
 -\tfrac{1}{30} g_{m_2 n_1} g_{m_1 n_2} R_{q_1 q_2 r_1 r_2} R_{r_1 r_2 p_1 p_2} R_{p_1 p_2 q_1 q_2} ,\\
\check{X}^6_{m_1 m_2 n_1 n_2} &= \tfrac23 R_{m_1 p_1 n_1 p_2} R_{m_2 q_1 p_1 q_2} R_{n_2 q_1 p_2 q_2}
+ \tfrac13  R_{m_1 p_2 n_1 p_1} R_{m_2 q_1 p_1 q_2} R_{n_2 q_1 p_2 q_2} \\&\ph{:=}		
+ \tfrac{1}{12} R_{m_1 m_2 p_1 p_2} R_{n_1 q_1 p_1 q_2} R_{n_2 q_1 p_2 q_2} 
+ \tfrac{1}{12} R_{n_1 n_2 p_1 p_2} R_{m_1 q_1 p_1 q_2} R_{m_2 q_1 p_2 q_2}\\&\ph{:=}
 -\tfrac{1}{10}  g_{m_2 n_1}  R_{n_2 r_1 q_1 q_2} R_{q_1 q_2 p_1 p_2} R_{p_1 p_2 m_1 r_1} 
 +\tfrac25 g_{m_2 n_1}  R_{n_2 q_1 r_1 q_2} R_{q_1 p_1 q_2 p_2} R_{p_1 m_1 p_2 r_1} 
  \\&\ph{:=}+\tfrac15 g_{m_2 n_1}  R_{n_2 q_1 m_1 q_2} R_{q_1 p_1 p_2 r_1} R_{q_2 p_1 p_2 r_1}\\&\ph{:=}
  +\tfrac{1}{120} g_{m_2 n_1} g_{m_1 n_2} R_{q_1 q_2 r_1 r_2} R_{r_1 r_2 p_1 p_2} R_{p_1 p_2 q_1 q_2} -\tfrac{1}{30} g_{m_2 n_1} g_{m_1 n_2}R_{q_1  r_1 q_2 r_2} R_{r_1 p_1 r_2 p_2} R_{p_1 q_1 p_2 q_2}.
\end{aligned}
\end{equation*}

The two-index $\check{W}^i$ are symmetric traceless:
\begin{equation*}
\begin{aligned}
\check{W}^1_{m_1 n_1 } &= R_{m_1 q_1 q_2 n_2} R_{q_2 n_2 p_1 p_2} R_{ p_1 p_2 n_1 q_1} -\tfrac{1}{7} g_{m_1 n_1} R_{q_1 q_2 r_1 r_2} R_{r_1 r_2 p_1 p_2} R_{ p_1 p_2 q_1 q_2},\\
\check{W}^2_{m_1 n_1} &= R_{m_1 q_2 q_1 n_2} R_{q_2 p_1 n_2 p_2} R_{ p_1 n_1 p_2 q_1} -\tfrac{1}{7} g_{m_1 n_1} R_{q_1 r_1 q_2 r_2} R_{r_1 p_1 r_2 p_2} R_{ p_1 q_1 p_2 q_2},\\
\check{W}^3_{m_1 n_1} &= R_{m_1 q_2 n_1 p_1} R_{q_2 n_2 q_1 q_2} R_{ p_1 n_2 q_1 q_2}.
\end{aligned}
\end{equation*}

We have two scalars $\check{S}^i$:
\begin{equation*}
\begin{aligned}
\check{S}^1 &= R_{m_1 m_2 n_1 n_2} R_{n_1 n_2 p_1 p_2}R_{ p_1 p_2 m_1 m_2},\\
\check{S}^2 &= R_{m_1 n_1 m_2 n_2} R_{n_1 p_1 n_2 p_2}R_{ p_1 m_1 p_2 m_2}.\\
\end{aligned}
\end{equation*}

We also need the six-index $\check{V}^i$, which have a set of five antisymmetric indices, vanish if the six indices are antisymmetrised, and are traceless:
\begin{equation*}
\begin{aligned}
\check{V}^1_{m_1  n_1 n_2 n_3 n_4 n_5} &= R_{p_1 m_1 n_2 n_3} R_{p_1 n_4 q_1 q_2} R_{n_5 n_1 q_1 q_2}\\
&\ph{:=}- \tfrac13  g_{m_1 n_2} R_{p_1 p_2 n_5 n_1} R_{q_1 q_2 n_3 n_4} R_{p_1 p_2 q_1 q_2}
+ \tfrac23   g_{m_1 n_2} R_{p_1 p_2 n_5 n_1} R_{q_1 q_2 p_1 n_3} R_{q_1 q_2 p_2 n_4},\\
\check{V}^2_{m_1  n_1 n_2 n_3 n_4 n_5} &=  R_{p_1 q_1 n_3 n_4} R_{m_1 q_2 p_1 n_2} R_{q_1 q_2 n_5 n_1}\\
 &\ph{:=}+ \tfrac13 g_{m_1 n_2} R_{p_1 p_2 n_5 n_1} R_{q_1 q_2 p_1 n_3} R_{q_1 q_2 p_2 n_4}
 - \tfrac23   g_{m_1 n_2} R_{q_1 q_2 n_5 n_1} R_{p_1 q_1 p_2 n_3} R_{p_2 q_2 p_1 n_4},\\
\end{aligned}
\end{equation*}

and three 4-forms $\check{T}^i$:
\begin{equation*}
\begin{aligned}
\check{T}^1_{n_1 n_2 n_3 n_4} &= R_{p_1 p_2 n_1 n_2} R_{q_1 q_2 n_3 n_4} R_{p_1 p_2 q_1 q_2},\\
\check{T}^2_{n_1 n_2 n_3 n_4} &= R_{p_1 p_2 n_1 n_2} R_{q_1 q_2 p_1 n_3} R_{q_1 q_2 p_2 n_4},\\
\check{T}^3_{n_1 n_2 n_3 n_4} &= R_{q_1 q_2 n_1 n_2} R_{p_1 q_1 p_2 n_3} R_{p_2 q_2 p_1 n_4}.\\
\end{aligned}
\end{equation*}

The eleven-dimensional terms thus decompose as

\begin{equation*}
\begin{aligned}
X^5_{m_1 m_2 n_1 n_2} &= 	\check{X}^5_{m_1 m_2 n_1 n_2} -\tfrac{8}{45} g_{m_2 n_1} \check{W}^1_{m_1 n_2} -\tfrac{1}{315}  g_{m_2 n_1} g_{m_1 n_2}\check{S}^1,\\
X^6_{m_1 m_2 n_1 n_2} &= 	\check{X}^6_{m_1 m_2 n_1 n_2} +\tfrac{2}{45} \check{W}^1_{m_1 n_2} -\tfrac{8}{45} \check{W}^2_{m_1 n_2} -\tfrac{4}{45} \check{W}^3_{m_1 n_2} \\&\ph{:=}+\tfrac{1}{1260} g_{m_2 n_1} g_{m_1 n_2} \check{S}^1-\tfrac{1}{315}  g_{m_2 n_1} g_{m_1 n_2}\check{S}^2,\\
V^2_{m_1  n_1 n_2 n_3 n_4 n_5} &= \check{V}^2_{m_1  n_1 n_2 n_3 n_4 n_5}
 -\tfrac{10}{21} g_{m_1 n_1}\check{T}^2_{n_2 n_3 n_4 n_5}  +\tfrac{8}{21}g_{m_1 n_1}\check{T}^3_{ n_2 n_3 n_4 n_5} ,\\
W^1_{m_1 n_1} &= \check{W}^1_{m_1 n_1} +\tfrac{4}{77} g_{m_1 n_1} \check{S}^1, \\
W^2_{m_1 n_1} &= \check{W}^2_{m_1 n_1} +\tfrac{4}{77} g_{m_1 n_1}  \check{S}^2, \\ W^3_{m_1 n_1} &= \check{W}^3_{m_1 n_1} ,\\
T^i_{n_1 n_2 n_3 n_4}  &= \check{T}^i_{n_1 n_2 n_3 n_4} , \qquad S^i = \check{S}^i .
\end{aligned}
\end{equation*}

\section{More general solution}
\label{app:non-min}

In section~\ref{sec:11d-r4-min} we were able to find a ``minimal'' solution to the M-theory Lichnerowicz, in the sense that we did not utilise all the possible terms listed in table~\ref{tab:appR4a} in the construction~\eqref{eq:Dmin} of the supersymmetry operator $D_a$.  Let us now consider adding the remaining terms of the table and see if we can find more solutions:
\begin{equation}
\begin{aligned}
\Dgen' :\, &S \rightarrow T^*\otimes S,\\
\Dgen'_a\varepsilon &= D_a \varepsilon + \sum_{i=1}^2 s_i \nabla_a (S^i \varepsilon )   
	+ \sum_{i=1}^3 w_i\Big(\tfrac{9}{10}\nabla^b ( W^i_{ab} \varepsilon )  - \tfrac{1}{10} \gamma_a{}^b\nabla^c (W^i_{cb} \varepsilon ) 
 +  \gamma^{b_1 b_2}  \nabla_{b_1}( W^i_{ab_2} \varepsilon ) \Big) \\
     &\ph{:=} + \sum_{i=1}^3 z_i\gamma^{c_1\dots c_4} \nabla^b( Z^i_{abc_1\dots c_4} \varepsilon ) \\
       &\ph{:=}+ \sum_{i=1}^3  u_i \Big(
      -\tfrac12\gamma^{b_2 b_3}\nabla^{b_1}(U^i_{ab_1b_2 b_3}  \varepsilon)
       + \tfrac{3}{10} \gamma^{b_2 b_3}\nabla^{b_1}(U^i_{b_1ab_2 b_3} \varepsilon )
      \\&\ph{:=}
       -\tfrac{2}{105} \gamma_a{}^{b_1b_2 b_3}\nabla^{c}(U^i_{cb_1b_2 b_3} \varepsilon) 
       +\tfrac{1}{21}\gamma^{b_1b_2 b_3b_4} \nabla_{b_4}( U^i_{ab_1b_2 b_3}\varepsilon  )
       \Big),
\end{aligned}
\end{equation}

and
\begin{equation}
\begin{aligned}
\tilde{D}' :\, &T^*\otimes S \rightarrow S,\\
\tilde{D}'^{a}\psi_a&= \tilde{D}^a \psi_a  + \sum_{i=1}^2 s_i\gamma^{ab}   S^i\nabla_a\psi_b   + \sum_{i=1}^3  w_i\gamma^{ac}W^i_{ab}\Big(   \nabla_c \psi^b  -  \nabla^b \psi_c\Big) \\
     &\ph{:=}+ \sum_{i=1}^3 z_i \gamma^{c_1\dots c_4}   Z^i_{abc_1\dots c_4}    \nabla^b\psi^a\\
     &\ph{:=}+ \sum_{i=1}^3 u_i\Big(  \gamma^{a c_3} U^i_{ac_1c_2 c_3}   \nabla^{c_1}\psi^{c_2}
     -\tfrac{1}{21}\gamma^{bc_1c_2 c_3} U^i_{ac_1c_2 c_3} \big( \nabla^{a}\psi_b - \nabla_b\psi^a \big) \Big).
\end{aligned}
\end{equation}

As previously mentioned, the $S^i$ and $ W^i$ terms do not actually contribute to anything at this level, so their coefficients will be unconstrained. We get that the 2-form in the M-theory Lichnerowicz vanishes if
\begin{equation}
\begin{aligned}
 z_1=-\tfrac{1}{14}u_2 + \tfrac{3}{28} z_2 +\tfrac{3}{2} z_3,
\end{aligned}
\end{equation}
while eliminating the 4-forms is solved by:
\begin{equation}
\begin{array}{lll}
 x_1 = -  y_2 - \tfrac{1}{36} u_1, \quad  
 &x_7 = 16 y_2 , \quad
 &t_1 = -\tfrac{1}{81} u_1 , \quad  \\
 x_2 = 16 y_2 +\tfrac29 u_1 ,  \quad 
 &x_8 = 16 y_2 ,   \quad
 &t_2 = \tfrac{2}{135} ( 216 y_2 - 3 x_5  +2 u_1) , \quad \\
 x_3 = -2  y_2 - \tfrac{1}{18} u_1 , \quad
 &y_1 = -\tfrac14 y_2 - \tfrac{1}{144} u_1 , \quad
 &t_3 = -\tfrac{4}{405} ( 648 y_2 - 9 x_5 +u_1 ) , \quad \\
 x_4 = -16 y_2 ,   \quad
 &u_2 = 0 , \quad
 &v_1 = -\tfrac{5}{36} u_1 ,  \quad\\
 x_6 = -4 x_5 +\tfrac49 u_1 ,  \quad
 &u_3 = -u_1 ,  \quad
 &v_2 = \tfrac{5}{72} ( -216 y_2 + 45 x_5 -5 u_1 )  , \\
\end{array}
\end{equation}
such that in total we have, in addition to $y_2$ and $x_5$, that $u_1,z_2,z_3, w_i, s_i$ remain undetermined. The resulting scalars in the Lichnerowicz are
\begin{equation}
\begin{aligned}
(\tilde{D}D\varepsilon)|_{\text{scalar}}&=-\tfrac14 \mathcal{R} 
+ \tfrac12 x_5 (  R_{46} -4   A_7)\\
 &\ph{:=}-\tfrac{1}{2} \tfrac{1}{12} y_2( 192  R_{41} +384 R_{42} +24 R_{43} +12  R_{44}-192 R_{45} - 384 A_7)\\
 &\ph{:=}+\tfrac12 \tfrac{1}{144}  u_1 (  - 2 R_{43} - R_{44}  +8 R_{45} +16 A_7 ),
\end{aligned}
\end{equation}
and the 8-form is
\begin{equation}
\begin{aligned}
(\tilde{D}D\varepsilon)|_{\text{8-form}}&=- \tfrac{1}{4} y_2(  -\tfrac14 R\cdot Y^1 +  R\cdot  Y^2) + \tfrac{1}{4} \tfrac{1}{144} u_1 R\cdot Y^1.
\end{aligned}
\end{equation}
The coefficients $s_i,w_i,z_i$ do not contribute at all, and so the corresponding terms in the operators are completely superfluous at this given order and given our  simplifications.

We see that the $u_1$ freedom corresponds to the remaining invariant mentioned in~\cite{Peeters:2000qj,deRoo:1992zp}, which here takes the form $I_{Y_1}+\tfrac{1}{24}(I_X-\tfrac18 I_Z)$. Differently from $I_X \pm \frac18 I_Y$ we do not expect this combination to lead to a full $N=2$ invariant. However, the emergence of an $N=1$ invariant in an eleven dimensional setting  suggests the interesting possibility of extending the  $\LL$ method to M-theory on manifolds with boundary, notably on the Horava-Witten interval.

\newpage


\end{document}